\documentclass[reprint,amsmath,amssymb,aps,showpacs,prx,longbibliography]{revtex4-1}
\usepackage[utf8]{inputenc}

\usepackage{graphicx}
\usepackage{dcolumn}% Align table columns on decimal point
\usepackage{bm}% bold math
\usepackage[colorlinks=true,allcolors=blue]{hyperref}% add hypertext capabilities
\usepackage{amsmath}
\usepackage{amssymb}
\usepackage{amsthm}
\usepackage{mathrsfs}

\newcommand{\tr}{\mbox{\textnormal{tr}}}

\renewcommand{\bf}[1]{\textbf{#1}}
\usepackage{bm}

\newcommand{\BZ}{\textnormal{BZ}}

\theoremstyle{definition}

\newtheorem{example}{Example}

\usepackage{url}

\usepackage{xcolor}

\begin{document}
\title{The product of two independent Su-Schrieffer-Heeger chains yields a two-dimensional Chern insulator}
%\title{The external tensor product of topological phases of free fermions: ``SSH times SSH equals Chern insulator''}

\author{Bruno Mera}
\email{bruno.mera@tecnico.ulisboa.pt}
\affiliation{Instituto de Telecomunica\c{c}\~oes, 1049-001 Lisboa, Portugal}
\affiliation{Departmento de F\'{i}sica, Instituto Superior T\'ecnico, Universidade de Lisboa, Av. Rovisco Pais, 1049-001 Lisboa, Portugal}
\affiliation{Departmento de Matem\'{a}tica, Instituto Superior T\'ecnico, Universidade de Lisboa, Av. Rovisco Pais, 1049-001 Lisboa, Portugal}

\date{\today}% It is always \today, today,

\begin{abstract}
We provide an extensive look at Bott periodicity in the context of complex gapped topological phases of free fermions. In doing so, we remark on the existence of a product structure in the set of inequivalent phases induced by the external tensor product of vector bundles -- a structure which has not yet been explored in condensed-matter literature. Bott periodicity appears in the form of a generalized Dirac monopole built out of a given phase, which is equivalent to the product of a Dirac monopole phase with that same given phase. The complex \emph{K}-theory cohomology ring is presented as a natural way to store the information of these phases, with a grading corresponding to the number of Clifford symmetries modulo $2$. The K\"unneth formula allows us to derive the result that, for band insulators, the Su-Schrieffer-Heeger (SSH) chain in one dimension allows one to generate the \emph{K}-cohomology of the $d$-dimensional Brillouin zone. In particular, we find that the product of two SSH chains in independent momentum directions yields a two-dimensional Chern insulator. The results obtained relate the associated topological phases of charge-conserving band insulators and their topological invariants in all spatial dimensions in a unified way.
\end{abstract}

\maketitle

%\tableofcontents

\section{Introduction}
\label{sec: introduction}
In Kitaev's seminal paper~\cite{kit:09}, gapped phases of free fermions, such as topological insulators and superconductors, were classified according to \emph{K}-theory, exhibiting Bott periodicity -- twofold in the complex case and eightfold in the real case. The impact of this result cannot be understated as it resulted in a paradigm shift. Not only did it allow physicists to comprehend new phases of matter which do not follow the traditional Landau-Ginzburg paradigm of symmetry breaking and local order parameters, bringing sophisticated mathematics into condensed matter, but also because it unified previously known results on topological phases, such as the integer quantum Hall effect~\cite{tknn:82}, the Majorana chain~\cite{kit:01} and the quantum spin Hall effect~\cite{kan:mel:05, kan:mel:05:2}, and, moreover, predicted phases that were not known to exist. Following Kitaev's pioneering work, other phases of matter were classified within the same spirit. More concretely, appropriate versions of \emph{K}-theory have been used to classify other phases of matter, to name a few, Floquet insulators~\cite{roy:har:17}, topological crystalline materials~\cite{shi:sat:gom:17,steh:18, cor:chap:19}, and also topological phases of non-Hermitian systems~\cite{gon:18}. There is even a monograph dedicated to complex topological insulators from the point of view of \emph{K}-theory of $C^*$-algebras, with emphasis in the bulk-to-boundary principle~\cite{pro:sch:16}.

An account of the mathematics of Bott periodicity has also been presented in condensed-matter literature. In  Ref.~\cite{stone:10}, for instance, they begin with a symmetry group, the unitary for the complex case and the real orthogonal for the real case, and by adding symmetry-breaking operators, they uncover the periodic pattern (modulo $2$ and modulo $8$, for the complex and real cases, respectively) of associated symmetric spaces in the stable limit. In this same reference, it is mentioned that the \emph{K}-theory groups have a product structure, and the authors wonder about the possible natural interpretation of the product in the context of gapped phases.

In the present paper, we will focus on the charge-conserving topological phases of gapped free fermions which are described by complex \emph{K}-theory. In this setup, we will provide an in-depth review of how the \emph{K}-theory  groups emerge in this setting, while giving emphasis to the reduced \emph{K}-groups, representing the non-trivial piece which forgets about the number of bands, and the product structure -- a novel ingredient brought from the mathematics of \emph{K}-theory. We will use three alternative equivalent descriptions of complex \emph{K}-theory, which can be advantageous from the physical point of view. Namely, Hamiltonians, (Fermi) projectors and vector bundles. The first two are the most familiar to condensed matter physics and make clear contact with physical systems. The last one is the classical mathematical point of view from Atiyah~\cite{ati:89}. The product structure is presented in detail. We review complex Bott periodicity, using a version of Bott's original construction which turns out to be equivalent, in \emph{K}-theory, to the (external tensor) product by a Dirac monopole defined over the two-sphere. Our main result is that by introducing the \emph{K}-cohomology ring, which contains information on Hamiltonians graded by the possible Clifford symmetries and also has a product, one can use the K\"unneth formula to prove that the \emph{K}-cohomology of the $d$-dimensional Brillouin zone is generated in terms of the well-known Su-Schrieffer-Heeger (SSH) chain. In particular, we find the particularly remarkable result that the product of two SSH chains is a Chern insulator, an example of which is the paradigmatic Haldane anomalous Hall insulator~\cite{hal:88}. 

The paper is organized as follows. Section~\ref{sec: topological phases of free fermions over X and the functor K} contains the review on complex \emph{K}-theory in the context of gapped phases of free fermions, while introducing the product structure. In Sec.~\ref{subsec: K^0(X) and topological phases without Clifford symmetries}, we focus on $K^0(X)$, which appears in the absence of any symmetries other than charge symmetry. In Sec.~\ref{subsec: K^-1(X) and topological phases with a single Clifford symmetry}, we focus on $K^{-1}(X)$, which appears when one imposes one Clifford symmetry, such as chiral symmetry. In Sec.~\ref{subsubsec: K-1 and suspensions}, we relate the $\widetilde{K}^{0}$ and $K^{-1}$ groups through suspensions. In Sec.~\ref{subsec: K^-2(X), tildeK^0(S^2X) and topological phases with two Clifford symmetries: Bott periodicity}, we deal with $K^{-2}(X)$ and Bott periodicity. Finally, in Sec.~\ref{sec: Kunneth formula and SSH times SSH equals Chern insulator}, we provide the definition of the \emph{K}-cohomology together with its ring structure, while presenting, in Sec.~\ref{subsec: ssh times ssh equals chern and k* bzd}, our main results that the \emph{K}-cohomology of the $d$-dimensional Brillouin zone is generated by SSH chains and SSH times SSH equals Chern insulator, and, in Sec.~\ref{subsec: relation to the topological invariant description}, the derived consequences for the associated topological invariants.

\section{Topological phases of free fermions over $X$ and the functor $K$}
\label{sec: topological phases of free fermions over X and the functor K}
In the following, we recall the arguments that relate complex, i.e. charge conserving, gapped phases of free fermions to \emph{K}-theory. We will put emphasis on three alternative equivalent descriptions of complex \emph{K}-theory: Hamiltonians, (Fermi) projectors, and vector bundles. The discussion is by no means complete and we omit details of known proofs. Therefore, we redirect the reader to standard references on \emph{K}-theory such as Refs.~\cite{ati:89, kar:08, hat:03}, and~\cite{par:08} for the connection with idemptotents and invertibles, which is quite natural in the physical context of free fermions. We will also review Bott periodicity in the present context, which can be seen as mapping a family of Hamiltonians over a space to a generalized Dirac monopole-like associated family over the Cartesian product of the original space by a two-sphere $S^2$, which is the same, up to deformation, to the homotopy equivalence presented by Bott in his original work from Refs.~\cite{bot:57, bot:59} -- see Eq.~\eqref{eq: generalized Dirac monopole}. In the process, we emphasize the fact that gapped topological phases of free fermions come equipped with a product, other than the direct sum, induced by the external tensor product of vector bundles, motivating the main results of the paper presented in Sec.~\ref{sec: Kunneth formula and SSH times SSH equals Chern insulator}. 
\subsection{$K^0(X)$ and topological phases without Clifford symmetries} 
\label{subsec: K^0(X) and topological phases without Clifford symmetries}
Suppose we are given a family of charge conserving free fermion Hamiltonians parametrized by a topological space $X$ which is compact, connected, and Hausdorff, namely,
\begin{align}
\label{eq: second quantized Hamiltonian}
\mathcal{H}(x)=\sum_{j,k=1}^{N}a^{\dagger}_j h_{jk}(x)a_k,\ x\in X,
\end{align}
where $\{a^{\dagger}_j\}_{j=1}^{N}$ denote fermion creation operators, and $H(x)=[h_{jk}(x)]_{1\leq j,k\leq N}$ is an Hermitian matrix whose entries are continuous functions. For convenience, we will also assume that $X$ is a pointed space, i.e., it comes equipped with a choice of a point $x_0\in X$. Here $X$ can be the $d$-dimensional Brillouin zone $\BZ^d$, see Sec.~\ref{sec: Kunneth formula and SSH times SSH equals Chern insulator}, but it can also be the space of parameters of the theory, such as the possible hopping amplitudes or fluxes thread through the system. In the case that $X=\BZ^d$, the fermionic creation and annihilation operators also depend on the momentum $\bf{k}\in \BZ^d$, but they are globally defined so the representation of Eq.~\eqref{eq: second quantized Hamiltonian} is an effective valid description. We remark that one can relax the compactness condition on $X$ to be locally compact, by taking suitable boundary conditions at ``infinity'' (one point compactification)-- compare the definitions below to those in Sec.~2.6 of Ref.~\cite{par:08}. See also Kitaev's discussion on continuous systems and Dirac operators in Ref.~\cite{kit:09}. 

The statement that the continuous family $\{\mathcal{H}(x)\}_{x\in X}$ is charge conserving means that the total charge
\begin{align}
Q=\sum_{j=1}^{N}a^\dagger_ja_j,
\end{align}
commutes with $\mathcal{H}(x)$ for all $x\in X$. From now on, we will identify $\{\mathcal{H}(x)\}_{x\in X}$ with $\{H(x)\}_{x\in X}$, since they are in one-to-one correspondence. Alternatively, we can just think of the continuous family $\{H(x)\}_{x\in X}$ as a continuous map $H:X\to \mbox{M}(N;\mathbb{C})$,
\begin{align}
H: X\ni x\mapsto H(x)\in \mbox{M}(N;\mathbb{C}),
\end{align}
where $\mbox{M}(N;\mathbb{C})$ denotes the set of $N\times N$ matrices with complex entries, and simply write it as $H$.

For the family $\{H(x)\}_{x\in X}$ or, equivalently, for $H$ to be \emph{admissible}, it must satisfy a gap condition. The gap condition adopted in Ref.~\cite{kit:09} is that the eigenvalues of $H(x)$ satisfy $\alpha \leq \lambda \leq \alpha^{-1}$, with $\alpha \in (0,1]$, for every $x\in X$. This condition assumes that the Fermi level $E_F$ of the system is set to zero. Within this set of admissible families we will impose an equivalence relation that identifies those families which belong to the same ``phase.'' 

The first classification principle is that of \emph{adiabatic continuity} or \emph{homotopy}. Namely, we say that two families $H_0:X\to \mbox{M}(N;\mathbb{C})$ and $H_1:X\to \mbox{M}(N;\mathbb{C})$ of admissible Hamiltonians are adiabatically connected or, equivalently, \emph{homotopic}, if there is a continuous path of admissible families that joins the two, i.e., if there is an admissible family $H:[0,1]\times X\to \mbox{M}(N;\mathbb{C})$ such that
\begin{align}
H(0,x)=H_0(x) \text{ and } H(1,x)=H_1(x) \text{ for all } x\in X.
\end{align}

The map $H:[0,1]\times X\to \mbox{M}(N;\mathbb{C})$ defines a homotopy between the two families, within the space of admissible families. Adiabatic connectivity defines an equivalence relation within the set of admissible families. By spectral flattening, one can show that every family has a representative with the property that
\begin{align}
H(x)^2=I_N, \text{ for every } x\in X,
\end{align}
where $I_N$ is the $N\times N$ identity matrix. The homotopy is explicitly given by $(1-t)H(x)+t \;\mbox{sgn}\left(H(x)\right)$, where $\mbox{sgn}(\cdot)$ is the sign function. The representative constructed only carries information about the \emph{Fermi projector}
\begin{align}
P(x)=\Theta(-H(x)),\ x\in X,
\end{align}
where $\Theta(\cdot)$ is the Heaviside step function. Indeed, $H(x)=I_N-2P(x)$. By this discussion, we see that, under homotopy, it is enough to begin with the set of continuous families of orthogonal projectors $\{P(x)\}_{x\in X}$, denoted also by the associated map $P:X\to\mbox{M}(N;\mathbb{C})$, and identify those that differ by homotopy. From now on we will think of $H$ also in terms of the associated family of orthogonal projectors $P$.

To be able to arrive at a notion of topological phases which is independent of $N$, we consider the inclusions $\mbox{M}(N;\mathbb{C})\subset \mbox{M}(N+1;\mathbb{C})$,
\begin{align}
\mbox{M}(N;\mathbb{C})\ni P\mapsto P\oplus 0 =\left[\begin{array}{cc}
P & 0\\
0 & 0
\end{array}\right]\in \mbox{M}(N+1;\mathbb{C}),
\end{align}
and consider the direct limit $\mbox{M}(\mathbb{C})=\lim\limits_{\longrightarrow} \mbox{M}(N;\mathbb{C})$, where we take the disjoint union $\coprod_{N\in\mathbb{N}} \mbox{M}(N;\mathbb{C})$ and quotient by the equivalence relation $P\sim P\oplus 0$, for every $P\in\mbox{M}(N;\mathbb{C})$ and $N\in\mathbb{N}$. In the direct limit, we can think of each finite dimensional matrix as a matrix which differs from the zero infinite matrix by a finite number of entries. The inclusion $\mbox{M}(N;\mathbb{C})\subset \mbox{M}(N+1;\mathbb{C})$ takes a family of orthogonal projectors $\{P(x)\}_{x\in X}$, with $P(x)\in \mbox{M}(N;\mathbb{C})$, and produces an orthogonal projector in $\mbox{M}(\mathbb{C})$. Observe however that this direct limit construction fails to preserve the gap condition if we take $\{H(x)\}_{x\in X}$ and identify it with a family with values in $\mbox{M}(\mathbb{C})$ by adding zeros. However, if $P\in \mbox{M}(N;\mathbb{C})$ is an orthogonal projector, then the matrix $I_{N+1}-2P\oplus 0\in \mbox{M}(N+1;\mathbb{C})$ still satisfies the gap condition. So, in terms of projectors, the construction above is perfectly legitimate in the sense that it preserves the gap condition. We could, however, observe that the gap condition implies that $H(x)$ is invertible and so is an element of $\mbox{GL}(N;\mathbb{C})$. The obvious direct limit construction would be to take the direct limit $\mbox{GL}(\mathbb{C})=\lim\limits_{\longrightarrow} \mbox{GL}(N;\mathbb{C})$, where every finite dimensional invertible matrix is thought of as an infinite matrix different from the infinite identity matrix from a finite number of entries. This is perfectly well defined, so we can think of the Hamiltonian as a continuous map $H:X\to \mbox{GL}(\mathbb{C})$.

From here on, we think of the family $\{P(x)\}_{x\in X}$ as a continuous map $P:X\to \mbox{M}(\mathbb{C})$, where $\mbox{M}(\mathbb{C})$ has the direct limit topology, i.e., a set is open if and only if the intersection with $\mbox{M}(N;\mathbb{C})$ is open for all $N\in\mathbb{N}$.

The resulting set of equivalence classes has the structure of what is called an \emph{Abelian monoid}, i.e., a set which has a  binary, closed, associative, and commutative operation with unit. This operation is direct sum and the zero element given by the zero matrix. A given continuous family of $N\times N$ orthogonal projectors $\{P(x)\}_{x\in X}$ has a topological vector bundle associated to it in a natural way, namely the vector bundle $\mbox{Im}\; P\to X$, whose fiber over $x\in X$ is the vector subspace $\mbox{Im}\; P(x)\subset\mathbb{C}^n$. Conversely, due to the result that every topological vector bundle over a compact Hausdorff space $X$ is isomorphic to a subbundle of the trivial bundle $X\times \mathbb{C}^N$ for some $N\in\mathbb{N}$ (see, for instance, Propositions~1.7.9 to 1.7.12 of Ref.~\cite{par:08}), every vector bundle, up to isomorphism, arises in this way. Moreover, if we have a vector bundle over $[0,1]\times X$, then we have a family of orthogonal projectors $\{P(t,x)\}_{(t,x)\in[0,1]}$ providing a homotopy between $\{P(0,x)\}_{x\in X}$ and $\{P(1,x)\}_{x\in X}$. The associated vector bundle $\mbox{Im}\;P$ restricted to $\{0\}\times X$ is isomorphic to $\mbox{Im}\;P$ restricted to $\{1\}\times X$. This result on homotopy invariance of vector bundles can be obtained using a version of the Tietze extension theorem and the existence of partitions of unity (see Lemma~1.4.3 of Atiyah's lectures\cite{ati:89}). Thus, we conclude that homotopic projectors give rise to isomorphic vector bundles. Moreover, if we have isomorphic vector bundles, then we can build a bundle over $[0,1]\times X$ with the two bundles corresponding to the restrictions to $\{0\}\times X$ and $\{1\}\times X$, respectively. To see this, one first realizes that if we have isomorphic vector bundles $E,F$, and without loss of generality we can assume $E,F\subset X\times \mathbb{C}^N$, associated to families of projectors $\{P^E(x)\}_{x\in X}$ and $\{P^F(x)\}_{x\in X}$, then there exists a continuous map $S:X\to \mbox{GL}(2N;\mathbb{C})$ with the property that $P^F\oplus 0_N=S(P^E\oplus 0_N)S^{-1}$ (see the proof of Proposition~1.7.6 of Ref.~\cite{par:08}). Then, one proves that projectors which satisfy this similarity relation are homotopic in double the dimension, using the homotopy
\begin{align}
T_t = (S \oplus 0_{2N})R_t (S^{-1} \oplus 0_{2N})R_{t}^{t},
\end{align}
where $R_t=\left[\begin{array}{cc}
\cos\left(\frac{\pi t}{2}\right) I_{2N} & -\sin\left(\frac{\pi t}{2}\right)I_{2N}\\
\sin\left(\frac{\pi t}{2}\right)I_{2N} & \cos\left(\frac{\pi t}{2}\right)I_{2N}
\end{array}\right]$, which interpolates between $I_{4N}$ and $S\oplus S^{-1}$, one has that $T_t(P^E\oplus 0_{N}\oplus 0_{2N})T_t^{-1}$ interpolates between $P^E\oplus 0_{3N}$ and $P^F\oplus 0_{3N}$. Thus, isomorphic vector bundles give rise to homotopic families of orthogonal projectors. This discussion shows that the isomorphism classes of vector bundles are in one-to-one correspondence with homotopy classes of families of orthogonal projectors. Additionally, the direct sum of orthogonal projectors yields precisely the Whitney direct sum of vector bundles and we can identify the set of equivalence classes of orthogonal projectors. In other words, what we have just built is the monoid of isomorphism classes of topological vector bundles over $X$, denoted $(\mbox{Vect}(X),\oplus)$. To turn this into an Abelian group, one considers the Grothendieck group completion of the monoid, which in terms of vector bundles is to consider $\mbox{Vect}(X)\times \mbox{Vect}(X)$, i.e., pairs of isomorphism classes of vector bundles $([E],[F])$ and quotient by the equivalence relation
\begin{align}
&([E],[F])\sim ([E'],[F']) \nonumber\\
&\Leftrightarrow\! \exists Q:\! E\oplus F'\oplus Q \cong E'\oplus F\oplus Q,
\end{align}
where $\cong$ denotes vector bundle isomorphism.
One often writes the equivalence class of $([E],[F])$ as a formal difference $[E]-[F]$. The addition of classes is given in terms of the Whitney sum as $([E_1]-[F_1])+([E_2]-[F_2])=[E_1\oplus E_2]-[F_1\oplus F_2]$. The zero element can be written as $[E]-[E]$ for any vector bundle $E$ and $-([E]-[F])=[F]-[E]$. The resulting Abelian group is known as $K^0(X)$ or simply $K(X)$ -- the complex topological \emph{K}-theory group of $X$.

Any element $[E]-[F]$ can be represented by $[E]-[\theta^n]$, where $\theta^n=X\times\mathbb{C}^n$ denotes the trivial bundle for some rank $n$. The reason is that, as explained above, due to the existence of partitions of unity over $X$ and the Tietze extension theorem, every vector bundle over $X$ is isomorphic to a subbundle of some trivial bundle. In terms of projectors, for any family $\{P(x)\}_{x\in X}$ of $N\times N$ orthogonal projectors, we can define an orthogonal complement family
\begin{align}
Q(x)=I_N-P(x),
\end{align}
so that $\mbox{Im}\; P\oplus \mbox{Im}\;Q=\theta^N$. If $[E]$ is an isomorphism class of vector bundles, we denote by $[E^\perp]$ the class obtained by considering the orthogonal complement family of projectors. Then
\begin{align}
[E]-[F]&=[E]-[F]+ [F^\perp]-[F^\perp]\nonumber\\
&=[E\oplus F^\perp] -[\theta^N],
\end{align}
for some natural number $N$. It is instructive to realize $K^0(X)$ as equivalence classes of pairs of families of projectors, in which addition is given by direct sum. As before, we denote the difference classes by $[\{P_1(x)\}_{x\in X}]-[\{P_2(x)\}_{x\in X}]$ or simply by $[P_1]-[P_2]$. It is important to note that, if $Q(x)=I_N-P(x)$, 
\begin{align}
P\oplus Q \text{ is homotopic to } (P+Q)\oplus 0_N=I_N\oplus 0_N,
\end{align}
and this gives us the equivalent statement in terms of pairs of equivalence classes of families of orthogonal projectors. Namely, there are trivial constant families $\{I_N\}_{x\in X}$, for $N\in\mathbb{N}$, and any $[P_1] -[P_2]$ can be written as
\begin{align}
&[P_1] -[P_2] +[I_N -P_2] -[I_N -P_2(x)]\nonumber\\
&=[P_1\oplus (I_N -P_2)] - [I_N], \text{ for some } N\in\mathbb{N}.
\end{align}
The equivalent statement in terms of Hamiltonians, is that we should think of topological phases as pairs of equivalence classes of families of Hamiltonians and we can take the second family to be a trivial family,
\begin{align}
\mathcal{H}=\sum_{j=1}^{N}\left(a_j^\dagger a_j-a_{N+j}^{\dagger}a_{N+j}\right), 
\end{align}
associated with the matrix $H=I_{N}\oplus (-I_N)$. We remark that the approach of taking equivalence classes of orthogonal projectors is equivalent to taking equivalence classes of idempotents, i.e., continuous maps $E:X\to \mbox{M}(\mathbb{C})$ satisfying $E(x)^2=E(x)$ for all $x\in X$. The reason is that any idempotent is homotopic, within the space of idempotents, to some family of orthogonal projectors. 

The assignment $X\mapsto K^0(X)$ is \emph{functorial} in the following sense. If we have a morphism of compact Hausdorff topological spaces, i.e., a continuous map $f:X\to Y$ between two such spaces $X$ and $Y$, then there is a group homomorphism $f^{*}:K^0(Y)\to K^0(X)$ in which 
\begin{align}
K^0(Y)\ni [E]-[F]\mapsto [f^*E] -[f^*F]\in K^0(X),
\end{align}
where $f^*E$ is the pullback bundle whose fiber at $x$ is the fiber of $E$ at $f(x)$. In terms of projectors it just means that given a family $\{P(y)\}_{y\in Y}$, we have a natural family over $X$ given by $\{P(f(x))\}_{x\in X}$. Obviously, if $\mbox{id}_X:X\to X$ is the identity map, the induced map in $\emph{K}$-theory is the identity map. Moreover, if two maps $f,g:X\to Y$ are homotopic, due to homotopy invariance of the induced maps on vector bundles by pullback, they induce the same maps $f^*,g^*:K^0(Y)\to K^0(X)$. Moreover, since pullbacks preserve direct sums, this induces a homomorphism of Abelian groups. In mathematical terms the assignment $X\mapsto K^0(X)$ is a contravariant functor from the category of compact Hausdorff topological spaces to that of Abelian groups. 

Taking the reference point $x_0\in X$, the natural inclusion $i:\{x_0\}\to X$ induces a map $i^*: K^0(X)\to K^0(\{x_0\})$. Now $K^0(\{x_0\})\cong \mathbb{Z}$, where the isomorphism is given by  $[E]-[F]\mapsto \dim E-\dim F$, or in terms of projectors, it is simply the difference between the ranks of the projectors. The kernel of $i^*$ is known as the reduced \emph{K}-theory group $\widetilde{K}^0(X)=\ker i^*$, also denoted by $\widetilde{K}(X)$. An element $[E]-[F]\in \widetilde{K}^0(X)$ has the property that $[i^*E]-[i^*F]=0$, in other words, the ranks of $E$ and $F$ have to be the same at $x_0$. We can then represent elements of $\widetilde{K}^0(X)$ in the form $[E]-[\theta^{n}]$, where $n=\dim E_{x_0}$. Another equivalent interpretation of $\widetilde{K}^0(X)$ is in terms of \emph{stable equivalence classes of vector bundles over} $X$. Namely, let $[E],[F]\in \mbox{Vect}(X)$ denote isomorphism classes of vector bundles over $X$, and introduce an equivalence relation given by
\begin{align}
[E]\sim_{s}[F] \text{ iff } E\oplus \theta^m\cong F\oplus \theta^n, \text{ for some } m,n.
\end{align}
The set of stable equivalence classes of vector bundles over $X$ is the quotient $\mathcal{E}\mathcal{U}(X)=\mbox{Vect}(X)/\sim_s$. Denote by $[E]_s$ the equivalence class of $[E]$ under the equivalence relation $\sim_s$. Then, if $[E]-[\theta^n]\in\widetilde{K}^0(X)$ we may map it to $[E]_s$. Observe that, in $\widetilde{K}^0(X)$, the equality $[E]-[\theta^n]=[F]-[\theta^m]$ implies there exists a vector bundle $Q$ such that
\begin{align}
E\oplus \theta^m\oplus Q\cong F\oplus \theta^n\oplus Q.
\end{align}
By noting that there exists $Q^{\perp}$ such that $Q\oplus Q^{\perp}\cong \theta^r$ for some $r$, the above condition is equivalent to $[E]_s=[F]_s$. By taking the direct sum of bundles in $\mathcal{E}\mathcal{U}(X)=\mbox{Vect}(X)/\sim_s$, the identification of $\widetilde{K}^0(X)$ with $\mathcal{E}\mathcal{U}(X)$ becomes a group homomorphism. The stable equivalence relation essentially \emph{forgets} about the dimensionality of the vector bundles, and it cares only about the \emph{non-trivial twist} of the bundle.

The \emph{K}-theory $K^0(X)$ is not only a group, but also a \emph{ring}, with the ring structure induced by tensor product of bundles. Namely, we can define
\begin{align}
&([E_1]-[F_1])([E_2]-[F_2])\nonumber\\
&=[E_1\otimes E_2\oplus F_1\otimes F_2] -[E_1\otimes F_2\oplus F_1\otimes E_2],
\end{align}
for bundles $E_1,E_2,F_1,F_2$ over $X$, and one can check that this is well-defined. Moreover, the pullback as defined before induces ring homomorphisms. By restriction, and because the kernel of a ring homomorphism is an ideal, we get a ring structure on the reduced \emph{K}-theory $\widetilde{K}^0(X)$. Thinking of the identification of $\widetilde{K}^0(X)$ with $\mathcal{E}\mathcal{U}(X)$, we would lake to make the latter into a ring with the same ring structure as the former. One would expect the product of stable isomorphism classes to be related to the tensor product. Indeed, taking $[E]-[\theta^n], [F]-[\theta^m]\in \widetilde{K}^0(X)$, we can write,
\begin{align}
&([E]-[\theta^n])([F]-[\theta^m])\nonumber\\
&=[E\oplus F\oplus \theta^{nm}] -[E\otimes \theta^m\oplus \theta^n\otimes F]\nonumber\\
&=[E\oplus F\oplus E^{\perp}\otimes \theta^m\oplus \theta^n\otimes F^{\perp}\oplus \theta^{mn}]\nonumber\\
&-[(E\oplus E^{\perp})\otimes \theta^{m} \oplus \theta^{n}\otimes (F\oplus F^{\perp})]\nonumber\\
&=[E\oplus F\oplus E^{\perp}\otimes \theta^m\oplus \theta^n\otimes F^{\perp}]-[\theta^{pm+nq-nm}], 
\end{align}
where $E^{\perp},F^{\perp}$ are orthogonal complement bundles such that $E\oplus E^{\perp}\cong \theta^{p}$ and $F\oplus F^{\perp}\cong \theta ^q$. So the good definition making $\mathcal{E}\mathcal{U}(X)\cong \widetilde{K}^0(X)$ into a ring isomorphism is to define
\begin{align}
[E]_s*[F]_s&=[E*F]_s\nonumber\\
&=[E\otimes F\oplus E^{\perp}\otimes \theta^m \oplus \theta^n\otimes F^{\perp}]_s,
\end{align}
where we defined the $*$ operation also on vector bundles by the the formula inside the bracket on the right-hand side.
Alternatively, we can think of \emph{stable equivalence classes of projectors} and \emph{stable equivalence classes of Hamiltonians}. That would mean that we would identify a class $[P]-[I_N]\in \widetilde{K}^0(X)$, with $N=\tr P(x_0)$ with a stable equivalence class in the following sense. Recall that $[P]$ is interpreted as a homotopy class of families of orthogonal projectors in $\mbox{M}(\mathbb{C})$. We then impose the further equivalence relation, 
\begin{align}
[P_1]\sim_s[P_2] \text{ iff } [P_1\oplus I_n]= [P_2\oplus I_m] \text{ for some } m,n, 
\end{align}
and denote the resulting equivalence class of $[P]$ in the quotient space by $[P]_s$.
Similarly, in terms of equivalence classes of Hamiltonians,
\begin{align}
&[H_1]\sim_s [H_2]\nonumber\\
&\text{ iff } [H_1\oplus \left( I_n\oplus \left(-I_n\right)\right)]= [H_2\oplus \left( I_m\oplus \left(-I_m\right)\right)]\nonumber,\\
&\text{ for some } m,n, 
\end{align}
and denote the resulting equivalence class of $[P]$ in the quotient space by $[H]_s$. It is not hard to see that the resulting sets are in bijection with $\mathcal{E}\mathcal{U}(X)$ or $\widetilde{K}^0(X)$, these bijections inducing group isomorphisms. Moreover, we can make it into ring isomorphisms by defining the product as follows. For stable equivalence classes of projectors,
\begin{align}
&[P_1]_s *[P_2]_s=[P_1*P_2]_s\nonumber\\
&= [P_1\otimes P_2 \oplus (I_{n_1}-P_1)\otimes I_{n_2}\oplus I_{n_1}\oplus (I_{n_2}-P_2)]_s,
\end{align}
where $n_i$, $i=1,2$, are the dimensions of the vector spaces where $P_1$ and $P_2$ act, so $I_{n_i}-P_i$, $i=1,2$, are the associated orthogonal projectors. For stable equivalence classes of Hamiltonians,
\begin{align}
&[H_1]_s *[H_2]_s= [H_1*H_2]_s\nonumber\\
&= [H_1\otimes H_2 \oplus (-H_1)\otimes \left(I_{n_2}\oplus \left(-I_{n_2}\right)\right)\nonumber\\
&\oplus \left(I_{n_1}\oplus \left(-I_{n_1}\right)\right)\oplus (-H_2)]_s,
\end{align}
where $n_i$ is the number of negative eigenvalues of $H_i$, $i=1,2$.
The above formula fixes, up to homotopy, what will be the product of two topological phases parametrized by $X$ with no Clifford symmetries. We will illustrate this in an example.

\begin{example}
\label{example:product of dirac monopoles}
This example will show that the product of Dirac monopoles is trivial. Let $X=S^2\subset \mathbb{R}^3$ and take 
\begin{align}
\label{eq: Dirac monopole}
H(x)=x^1\sigma_1+x^2\sigma_2+x^3\sigma_3=\vec{x}\cdot\vec{\sigma},
\end{align}
where $\vec{\sigma}=(\sigma_1,\sigma_2,\sigma_3)$ are the Pauli matrices. The associated projector,
\begin{align}
P(x)=\frac{1}{2}\left(1+ H(x)\right),
\end{align}
defines an identification $S^2$ with the rank $1$ orthogonal projectors in $\mathbb{C}^2$ or, equivalently, with the space of one-dimensional subspaces of $\mathbb{C}^2$, i.e., $\mathbb{C}P^1$. The associated bundle $\mbox{Im}\; P\to S^2$ is then identified with the tautological bundle over $\mathbb{C}P^1$, denoted $\mathcal{L}\to \mathbb{C}P^1$, whose fiber over a one dimensional subspace is the subspace itself. In physical terms, the associated bundle is seen as the charge $-1$ Dirac monopole bundle, since the associated Berry curvature can be identified with a magnetic field of a magnetic monopole sitting at the origin of $\mathbb{R}^3$, of topological charge $-1$. This charge is most simply the first Chern number of the bundle $\mbox{Im}\;P\to S^2$. Let us take the external tensor product Hamiltonian:
\begin{align}
\left(H*H\right)(x)&=H(x)\otimes H(x)\oplus \left(-H(x)\right)\otimes\sigma_3\nonumber\\
&\oplus \sigma_3\otimes\left(-H(x)\right).
\end{align}
The associated bundle of positive eigenvalues is isomorphic to
\begin{align}
\mathcal{L}\otimes \mathcal{L}\oplus \mathcal{L}^{\perp}\otimes\theta^1\oplus \theta^1\otimes \mathcal{L}^{\perp},
\end{align}
which is seen to be isomorphic to a trivial bundle because the associated Chern number is $0$. In terms of \emph{K}-theory, this can be seen due to the fact that $S^2$ can be written as the union of two disks which are contractible and, thus, the product in reduced $K$-theory is trivial, see Example $2.13.$ of Ref.~\cite{hat:03}. 
\end{example}

For any $X,Y$, we have the \emph{external tensor product} given by
\begin{align}
\mu:\; & K^0(X)\otimes K^0(Y)\longrightarrow K^0(X\times Y)\nonumber\\
  & a\otimes b\longmapsto  p_1^*a \; p_2^*b,
\end{align}
where $p_1:X\times Y\to X$ and $p_2:X\times Y\to Y$ are the canonical projections. Restriction to $\widetilde{K}^0(X)\otimes \widetilde{K}^0(Y)$ also yields a map 
\begin{align}
\mu: \widetilde{K}^0(X)\otimes \widetilde{K}^0(Y)\longrightarrow \widetilde{K}^0(X\times Y),
\end{align}
in fact, it yields a map to $\widetilde{K}(X\wedge Y)$, known as the \emph{smash product} of $X$ and $Y$, which is obtained by taking the Cartesian product $X\times Y$ and collapsing the subspace corresponding to the \emph{wedge sum} $X\vee Y=\{x_0\}\times Y\sqcup X\times \{y_0\}$. The reason being that $a$ is $0$ as an element of $K(\{x_0\})$ and its pullback by $p_1$ is zero over $\{x_0\}\times Y$ (meaning it maps to zero under the induced map by the inclusion $\{x_0\}\times Y\hookrightarrow X\times Y$) and, similarly, the pullback by $p_2$ of $b$ is zero over $X\times\{y_0\}$. Therefore it maps to zero under the induced map $\widetilde{K}(X\times Y)\to \widetilde{K}(X\vee Y)$ and defines an element of $\widetilde{K}(X\wedge Y)$. See, for example, Ref.~\cite{hat:03} for a detailed proof.
Observe that, as a consequence, we can also define an \emph{external tensor product} of stable equivalence classes of vector bundles by the formula
\begin{align}
\mu: & \mathcal{E}\mathcal{U}(X)\otimes \mathcal{E}\mathcal{U}(Y)\longrightarrow \mathcal{E}\mathcal{U}(X\wedge Y)\nonumber\\
  & [E]_s\otimes [F]_s\longmapsto  [p_1^*E]_s *[p_2^*F]_s,
\end{align}
and similarly for stable equivalence classes of projectors and Hamiltonians.
The famous theorem of Bott periodicity stems from the fact that $\mu: \widetilde{K}^0(X)\otimes \widetilde{K}^0(S^2)\to \widetilde{K}^0(X\wedge S^2)$ is an isomorphism of Abelian groups. More concretely, if we take $b=[\mathcal{L}]-[\theta^1]\in \widetilde{K}^0(S^2)$, where $\mathcal{L}\to S^2$ is the tautological line bundle, the map $\alpha: a\mapsto a*b\in \widetilde{K}^0(X\wedge S^2)$, for $a\in \widetilde{K}^0(X)$, is an isomorphism of Abelian groups. Observe that in terms of stable equivalence classes of vectors bundles, this corresponds to
\begin{align}
&\mathcal{E}\mathcal{U}(X)\ni [E]_s\mapsto  [p_1^*E *p_2^*\mathcal{L}]_s\in \mathcal{E}\mathcal{U}(X\wedge S^2),
\end{align}
where $[p_1^*E*p_2^*\mathcal{L}]_s=[p_1^*E\otimes p_2^*\mathcal{L}\oplus p_1^*E^\perp\otimes \theta^1\oplus \theta^n\otimes p_2^*\mathcal{L}^{\perp}]_s$. For more details on the isomorphism, see Sec.~\ref{subsec: K^-2(X), tildeK^0(S^2X) and topological phases with two Clifford symmetries: Bott periodicity}.
\subsection{$K^{-1}(X)$ and topological phases with a single Clifford symmetry} 
\label{subsec: K^-1(X) and topological phases with a single Clifford symmetry}
Within the discussion above, we did not address the possibility of having additional generic symmetries. Here the word generic means a symmetry, such as chiral symmetry, that has an implementation at the level of the single particle sector of the theory in terms of a Clifford algebra generator. When we do this, we obtain other $K$-groups. In fact, due to Bott-periodicity, in the complex case, there are only two such groups $K^0(X)$ and $K^1(X)$. Let us describe the latter. Now we consider that our admissible families $\{H(x)\}_{x\in X}$  are $N\times N$ Hermitian gapped matrices satisfying
\begin{align}
H(x)\Gamma=-\Gamma H(x), \text{ for all } x\in X,
\end{align}
for $\Gamma$ a constant matrix with $\Gamma^2=I_N$. Because of this if $\lambda$ is an eigenvalue of $H(x)$ and $v$ an associated eigenvector, then $\Gamma v$ is an eigenvector with eigenvalue $-\lambda$. Since $\Gamma^2=I_N$, $\Gamma$ provides an isomorphism between the positive energy eigenspaces and the negative energy eigenspaces. Because of the gap condition, $N$ must be even. We now replace the $N$ by $2N$. We can choose a basis where 
\begin{align}
\Gamma=\left[\begin{array}{cc}
I_N & 0\\
0 & -I_N
\end{array}\right],
\end{align}
and this implies that our allowed families must satisfy
\begin{align}
H(x)=\left[\begin{array}{cc}
0 & U^\dagger(x)\\
U(x) & 0
\end{array}\right],
\end{align}
where $U(x)\in \mbox{U}(N)$ is unitary for each $x\in X$. To get rid of the dimension label, the matrix $\Gamma$ must be rescaled in the appropriate way, one more positive eigenvalue and one more negative eigenvalue at a time. The $N\times N$ unitary matrix $U(x)$ naturally fits in a $(2N+2)\times (2N+2)$ matrix $H(x)$ by taking the inclusion $\mbox{U}(N)\subset \mbox{U}(N+1)$
\begin{align}
\mbox{U}(N)\ni U\mapsto U\oplus 1=\left[\begin{array}{cc}
U & 0\\
0 & 1
\end{array}\right]\in \mbox{U}(N+1).
\end{align}
Taking the direct limit, we get continuous maps $U: X\to \mbox{U}$, where $\mbox{U}=\lim\limits_{\longrightarrow}\mbox{U}(N)$ with the direct limit topology. The homotopy classes of such maps form an Abelian group $K^{-1}(X)$ under the usual matrix multiplication. In a similar fashion to what happened with $K^{0}(X)$, continuous maps $f:X\to Y$, induce group homomorphisms $f^*:K^{-1}(Y)\to K^{-1}(X)$ going in the opposite direction. The minus in the notation for $K^{-1}(X)$ is justified, mathematically, with the fact that the Abelian groups $K^{-n}(X)$, $n\in\mathbb{Z}$, form a \emph{generalized cohomology theory}, and this will be explored in Sec.~\ref{sec: Kunneth formula and SSH times SSH equals Chern insulator}.
\subsubsection{$K^{-1}(X)$ and $\widetilde{K}^0(SX)$}
\label{subsubsec: K-1 and suspensions}
The group $K^{-1}(X)$ as defined above is isomorphic to $\widetilde{K}^0(SX)$, where $SX$ denotes the \emph{suspension of} $X$. Recall that the suspension of $X$ is the topological space obtained from $X\times [0,1]$ by collapsing $X\times \{0\}$ and $X\times \{1\}$ to a point. We will denote by $q:X\times [0,1]\to SX$ the quotient map and the equivalence classes in the quotient by $[(t,x)]=q(t,x)$, where $t\in [0,1]$, $x\in X$. Another important space is the \emph{reduced suspension} $\Sigma X$ which is the quotient of $SX$ by further collapsing the line $\{x_0\}\times [0,1]$. Note that this space is homeomorphic to $X\wedge S^1$ and also, since $\{x_0\}\times [0,1]$ is contractible within $SX$, it has the same homotopy type of $SX$. As a consequence $\widetilde{K}^0(SX)\cong \widetilde{K}^0(\Sigma X)\cong \widetilde{K}^0(X\wedge S^1)$. In fact, the usual definition of the higher order \emph{K}-groups is, see Refs.~\cite{ati:89, hat:03}, $\widetilde{K}^{-n}(X)=\widetilde{K}(\Sigma^nX)$, for $n\geq 0$ (and defined using Bott periodicity for negative $n$). Since $K^{-1}(\{x_0\})$ is trivial, because the unitary groups are path connected, we have that $K^{-1}(X)\cong \widetilde{K}^{-1}(X)$, the latter defined as the kernel of the group homomorphism induced by the inclusion $i:\{x_0\}\hookrightarrow X$, and the isomorphism we are building is equivalent to $\widetilde{K}^{-1}(X)\cong \widetilde{K}^{0}(SX)$. We will now provide the explicit isomorphism $K^{-1}(X)\cong \widetilde{K}^0(SX)$, within the definitions coming from phases of gapped free fermions.

From a map $U:X\to \mbox{U}(X)$ we can construct a vector bundle over $SX$ by taking the trivial bundle over the \emph{cones} $C_{-}$ and $C_{+}$, corresponding to the projections, respectively, of $X\times[0,1/2]$ and $X\times [1/2,1]$ onto the quotient, and gluing them together in the overlap $X\times\{1/2\}\cong X$ through the ``clutching function'' $U$. In terms of matrices this can be achieved as follows. Define
\begin{align}
\label{eq: suspension hamiltonian}
&H(t,x)=\cos(\pi t)\Gamma +\sin(\pi t) H(x)\nonumber\\
&=\left[\begin{array}{cc}
\cos(\pi t)I_N & \sin(\pi t) U^\dagger(x)\\
\sin(\pi t)U(x) & -\cos(\pi t) I_N
\end{array}\right], \text{ for } (t,x)\in [0,1]\times X.
\end{align}
Observe that $H(0,x)=H(1,x)=\Gamma$, so that $H(x,t)$ defines a family over the suspension $SX$, and $H(1/2,x)=H(x)$. The associated projector,
\begin{align}
P(t,x)=\frac{I_{2N}-H(t,x)}{2},
\end{align}
defines a vector bundle $\mbox{Im}\; P \to SX$ with the desired properties. To see that this is the case, observe that the columns of the matrix
\begin{align}
&v_{-}(t,x)=\frac{1}{\sqrt{1+\tan^2\left(\frac{\pi t}{2}\right)}}\left[\begin{array}{c}
I_N\\
\tan\left(\frac{\pi t}{2}\right) U(x)
\end{array}\right],\nonumber\\
&t\in [0,1), x\in X,
\end{align}
form a basis for the eigenspace of $H(t,x)$ of energy $1$ for every $t\in[0,1)$, $x\in X$. In the same way,
\begin{align}
&v_{+}(t,x)=\frac{1}{\sqrt{1+\cot^2\left(\frac{\pi t}{2}\right)}}\left[\begin{array}{c}
\cot\left(\frac{\pi t}{2}\right) U^\dagger(x)\\
I_N
\end{array}\right],\nonumber\\
&t\in (0,1], x\in X,
\end{align}
When $t=1/2$ and $x\in X$, we have
\begin{align}
v_{-}(1/2,x)=\left[\begin{array}{c}
I_N\\
U(x)
\end{array}\right] \text{ and }
v_{+}(1/2,x)=\left[\begin{array}{c}
U^{\dagger}(x)\\
I_N
\end{array}\right],
\end{align}
so we have a transition function $g_{+-}(x)=U(x)$ at the equator of the suspension. This construction provides the desired group isomorphism $K^{-1}(X)\cong \widetilde{K}^0(SX)$, namely, the assignment
\begin{align}
K^{-1}(X)\ni[U]\mapsto [P]-[I_N]\in \widetilde{K}^{0}(SX)\subset K^0(SX),
\end{align}	
where $N=\tr\; P(t,x)$, for all $t\in [0,1]$, $x\in X$. It is not hard to see that the map is injective, as different homotopy classes of maps $x\mapsto U(x)\in \mbox{U}$ can not be deformed continuously into each other so the resulting bundles can not be isomorphic. To see that this is indeed an isomorphism, we need to check that any element $\widetilde{K}^0(SX)$ occurs in this way. Take a continuous family of orthogonal projectors of rank $N$, $\{P(t,x)\}_{(t,x)\in I\times X}$, where $P(t,x)\in \mbox{M}(M;\mathbb{C})$, where, without loss of generality, we can assume $M>N$, as we will eventually take it as a family with values in $\mbox{M}(\mathbb{C})$. It defines a family over $SX$ if and only if $P(0,x)$ and $P(1,x)$ are both independent of $x\in X$. In this case we have a well-defined class $[P]-[I_N]\in \widetilde{K}^{0}(SX)$. Since $C_{-}$ and $C_{+}$ are contractible in $SX$, it means that the restriction of $\{P(t,x)\}_{(t,x)\in I\times X}$ to these subspaces provide homotopies to constant orthogonal projectors. Define $P(x)=P(1/2,x)$, $x\in X$. Then $P(x)$ is homotopic to a constant orthogonal projector, through $\{P(t,x)\}_{[(t,x)]\in C_{-}}$ and through $\{P(t,x)\}_{[(t,x)]\in C_{+}}$. Equivalently, the associated bundles $E_{\pm}:=\mbox{Im}\;P|_{C_{\pm}}$ are trivializable, trivializations which can be chosen to be unitary. The relation between them at the overlap $C_{+}\cap C_{-}=\{1/2\}\times X\cong X$ is provided by a unitary matrix $U:X\to \mbox{U}(N)$. Up to homotopy, we can always reconstruct $\{P(t,x)\}_{[(t,x)]\in SX}$ from the homotopy class of $U:x\mapsto U(x)\in \mbox{U}(N)$. From $U$, build the Hamiltonian
\begin{align}
H(t,x)=\cos(\pi t) \Gamma +\sin(\pi t) H(x),
\end{align}
with 
\begin{align}
H(x)=\left[\begin{array}{cc}
0 & U^\dagger(x)\\
U(x) & 0
\end{array}\right],
\end{align}
which concludes the proof. It is useful to extend the map to a loop of unitaries. Observe that at $t=1$, $H(t,x)=-\Gamma$. We can extend the map by declaring that from $1\leq t\leq 2$, we have
\begin{align}
H(t,x)=\cos(\pi t) \Gamma +\sin(\pi t)H_0,
\end{align}
with
\begin{align}
H_0=\left[\begin{array}{cc}
0 & I_N\\
I_N & 0
\end{array}\right].
\end{align}
The resulting map satisfies $H(0,x)=\Gamma=H(2,x)$ and, thus, defines a loop of Hamiltonians with no Clifford symmetries. The map we have just built is essentially $\lambda: \mbox{U}(N)\to \Omega G_N$, as described in Ref.~\cite{bot:59} where $G_{N}=\mbox{U}(2N)/\left(\mbox{U}(N)\times\mbox{U}(N)\right)$ is the \emph{Grassmannian} of $N$-planes in $\mathbb{C}^{2N}$ which can be identified with the orthogonal projectors of rank $N$ is $\mathbb{C}^{2N}$, or the set of Hermitian matrices in $\mathbb{C}^{2N}$ satisfying $H^2=I_N$ and having $N$ negative eigenvalues. The space $\Omega X$ is the \emph{loop space} of $X$, which is the space of based loops in $X$.

\subsection{$K^{-2}(X)$, $\widetilde{K}^0(S^2X)$ and topological phases with two Clifford symmetries: Bott periodicity}
\label{subsec: K^-2(X), tildeK^0(S^2X) and topological phases with two Clifford symmetries: Bott periodicity}
Next, we consider adding another Clifford symmetry $\Gamma_2$ and see that it reproduces the group $K^0(X)$. We let $\Gamma_1=\Gamma$ be defined as before and pick a fixed choice of $\Gamma_2$ satisfying
\begin{align}
\Gamma_i\Gamma_j+\Gamma_j\Gamma_i=2\delta_{ij}I_{2N},\ i,j=1,2.
\end{align}
One such choice is given by
\begin{align}
\Gamma_2=\left[\begin{array}{cc}
0_{N} & -i I_{N}\\
i I_{N} & 0_{N}
\end{array}\right].
\end{align}
Now we look for continuous families of $N\times N$ Hermitian matrices $\{H(x)\}_{x\in X}$ such that
\begin{align}
&H(x)^2=I_{2N} \text{ and } H(x)\Gamma_i +\Gamma_iH(x)=0,\nonumber\\
& \text{ for } i=1,2 \text{ and for all } x\in X. 
\end{align}
It is not hard to show that these matrices have the form
\begin{align}
\label{Eq:Hamiltonian 2-Clifford symmetric}
H(x)=\left[\begin{array}{cc}
0_{N} &  h(x)\\
h(x) & 0_{N}
\end{array}\right],
\end{align}
where $h(x)$ is an $N\times N$ Hermitian matrix and squares to the identity. It is clear that these families are completely determined by the $N\times N$ blocks $h(x)$, which in turn are determined by the orthogonal projector $p(x)=\Theta(-h(x))$. To get rid of the dependence on $N$, we must rescale both matrices $\Gamma_1$ and $\Gamma_2$ appropriately. The replacement $N\to N+1$ in these matrices is naturally accompanied by the inclusion
\begin{align}
h(x)\mapsto h(x)\oplus 1,
\end{align}
which in terms of the projector $p(x)$ means
\begin{align}
p(x)\mapsto p(x)\oplus 0.
\end{align}
Taking the direct limit and imposing the homotopy equivalence relation, we again obtain the monoid $(\mbox{Vect}(X),\oplus)$ and the Grothendieck group completion retrieves $K^0(X)$. Hence, what we would logically call $K^{-2}(X)$ is naturally isomorphic to $K^0(X)$. This is a manifestation of \emph{Bott periodicity} in physics. In Appendix~\ref{sec:2-fold bott periodicity}, we look at it from a different perspective again using the suspension construction, and show how the Dirac monopole or, equivalently, the tautological bundle over the sphere $\mathcal{L}\to S^2$ plays an important role in it. In particular, what we show there is that $\widetilde{K}^{-2}(X)$ which, under the above definition, is naturally identified with $\widetilde{K}^{0}(X)$, is isomorphic to $\widetilde{K}^{0}(X\wedge S^2)=\widetilde{K}^{0}(\Sigma^2X)$, which is the usual definition of $\widetilde{K}^{-2}(X)$. The isomorphism is given by taking the external tensor product with a Dirac monopole.

This concludes our digression through complex \emph{K}-theory and Bott periodicity in the context of gapped phases of free fermions. 
\section{K\"unneth formula and ``SSH times SSH equals Chern insulator''}
\label{sec: Kunneth formula and SSH times SSH equals Chern insulator}
From the Abelian groups $K^0(X)$ and $K^{-1}(X)$ one can build a graded group 
\begin{align}
K^{*}(X)&=K^0(X)\oplus K^1(X)\nonumber\\
&\cong K(\{x_0\})\oplus \widetilde{K}^{0}(X)\oplus \widetilde{K}^{-1}(X),
\end{align} 
known as the \emph{K-cohomology group of} $X$. Observe that this group stores information of complex gapped topological phases of free fermions with an arbitrary number of Clifford symmetries, so it is natural to consider it as a whole. Note that the grading is precisely given by the number of Clifford symmetries $\mod 2$.

If we take the usual definition $\widetilde{K}^{-i}(X)=\widetilde{K}^0(X\wedge S^i)$, we have a product 
\begin{align}
\label{eq: product in k-cohomology}
\mu: \widetilde{K}^0(X\wedge S^i)\otimes \widetilde{K}^0(Y\wedge S^j)\to &\widetilde{K}^0(X\wedge S^i \wedge X\wedge S^j)\nonumber\\
&= \widetilde{K}^0\left((X\wedge Y)\wedge S^{i+j}\right),
\end{align}
because for any compact Hausdorff spaces $X,Y,Z$, we have that $X\wedge (Y \wedge Z)\cong (X\wedge Y)\wedge Z$ and $X\wedge Y\cong Y\wedge X$, and $S^i\wedge S^j\cong S^{i+j}$ (see Ref.~\cite{ati:89} or Ref.~\cite{hat:03}, for instance). One uses the Bott class to provide isomorphisms $\widetilde{K}^{i}(X)\cong \widetilde{K}^{i+2}(X)$, for every $i$. Finally, to extend the product to the unreduced \emph{K}-groups~\cite{hat:03}, one can use the fact that $K^{-i}(X)=\widetilde{K}^{-i}(X_{+})$, with $X_{+}=X\sqcup\text{pt}$, where $\text{pt}$ denotes a point, and $i=0,1$. Indeed, $\widetilde{K}^{0}(X_{+})=K^0(X)$ and, since $X_{+}\wedge S^1\cong(X\wedge S^1)\vee S^1$, we have that $\widetilde{K}^{-1}(X_{+})=\widetilde{K}(X_{+}\wedge S^1)=\widetilde{K}^{-1}(X)\oplus \widetilde{K}^0(S^1)=\widetilde{K}^{-1}(X)\oplus \widetilde{K}^0(S^1)=\widetilde{K}^{-1}(X)$, because vector bundles over the $S^1$ are trivializable (since the general linear group is path connected) and because~\cite{hat:03} $\widetilde{K}^0(X\vee Y)\cong \widetilde{K}^0(X)\oplus \widetilde{K}^0(Y)$, for any compact Hausdorff spaces $X,Y$. Using the maps induced by the diagonal map $\Delta: X\hookrightarrow X\times X$, given by  $x\mapsto (x,x)$, for all $x\in X$, one obtains a $\mathbb{Z}_2$-graded product defining a ring structure on $K^{*}(X)$, which can be shown to satisfy $ab=(-1)^{ij}ba$, for $a\in K^{i}(X), b\in K^{j}(X)$ and $i,j\in \{0,1\}$(because exchanging the two factors involves a permutation of $S^i$ and $S^j$ factors in $S^i\wedge S^j$, which in \emph{K}-theory yields the sign of the permutation-- see, for instance, Lemma~2.4.11 of Atiyah's lectures~\cite{ati:89}). 
One can show that the product $\mu$ (for spaces of finite type) defines a ring isomorphism:
\begin{align}
\label{eq:kunneth}
K^*(X\times Y)\cong K^*(X)\otimes K^{*}(Y),
\end{align}
known as the K\"unneth theorem, see Ref.~\cite{par:08}. 

Before proceeding, we would like to remark that the K\"{u}nneth theorem of Eq.~\eqref{eq:kunneth} is, in its origin, different from the K\"unneth theorem in ordinary cohomology. The K\"unneth formula in complex \emph{K}-theory is of a different nature as the objects considered are formal differences of isomorphism classes of vector bundles over a given Hausdorff compact topological space and not closed differential forms over a smooth manifold, nor Abelian \v{C}ech cocyles. In particular, the K\"unneth formula for ordinary cohomology appears in previous works in condensed-matter literature and, more precisely, in research works on topological phases, for example in Ref.~\cite{ann:kim:kim:yang:18}, within the context of Eq.~(41) and the discussion below. The K\"unneth formula there refers to ordinary cohomology, rather than the generalized cohomology associated with complex \emph{K}-theory.
\subsection{``SSH times SSH equals Chern insulator'' and $K^*(\BZ^d)=\Lambda(\textnormal{SSH}_1,...,\textnormal{SSH}_d)$}
\label{subsec: ssh times ssh equals chern and k* bzd}
As a consequence of the K\"unneth theorem, we have the following main result of our paper relating band insulators in $1$ and $2$ dimensions with charge symmetry, and more generally relating band insulators in $1$ dimension and those in $d$ dimensions. 
\begin{example}
\label{example:BZ2 and BZ1, BZd}
This example will show that $K^*(\BZ^2)=\Lambda(\text{SSH}_1,\text{SSH}_2)$ and, more generally, $K^*(\BZ^d)=\Lambda(\text{SSH}_1,...,\text{SSH}_d)$, where $\text{SSH}_i$, $i=1,...,d$ are SSH chains in the available independent momentum directions. If we observe that the Brillouin zone in $d$ dimensions is, topologically, a torus, i.e., $\BZ^d=T^d=S^1\times ...\times S^1$, we only need to look at $K^{*}(\BZ^1)=K^{*}(S^1)$. We have that
\begin{align}
K^{*}(S^1)&=K^0(S^1)\oplus K^1(S^1)=K^0(\{x_0\})\oplus \widetilde{K}^{-1}(S^1)\nonumber\\
&=K^0(\{x_0\})\oplus \widetilde{K}^0(S^2)= \mathbb{Z}\oplus \mathbb{Z}\;b,
\end{align}
where we noted that $\widetilde{K}^{-1}(S^1)=\widetilde{K}^{0}(S^2)\cong\mathbb{Z}$, generated by the Bott class $b=[\mathcal{L}]-[1]$. We proceed to describe multiplication in $K^*(S^1)$. The multiplication of elements of $K^0(S^1)=K^0(\{x_0\})=\mathbb{Z}$ is just the usual multiplication in $\mathbb{Z}$; the multiplication of elements of $K^0(S^1)=\mathbb{Z}$ by elements of $K^{-1}(S^1)=\mathbb{Z}\;b$ induces simple scalar multiplication $m\otimes nb\mapsto mn b\in K^{-1}(S^1)$, for $m,n\in\mathbb{Z}$, and, since $\widetilde{K}^0(S^1)=0$, multiplication of elements $\widetilde{K}^{-1}(S^1)$ yields zero. One concludes that
\begin{align}
K^*(\BZ^1)\cong \mathbb{Z}[b]/b^2\cong\Lambda(b),
\end{align}
as a $\mathbb{Z}_2$-graded ring, where $\mathbb{Z}[b]/b^2$ corresponds to polynomials in $b$ modded out by the relation $b^2=0$ and $\Lambda(b)$ denotes the exterior algebra generated by $b$. As a result, by the K\"unneth theorem,
\begin{align}
K^*(\BZ^2)\cong K^*(\BZ^1)\otimes K^*(\BZ^1)\cong \Lambda(b_1,b_2),
\end{align}
where $\Lambda(b_1,b_2)$ denotes the exterior algebra in two generators $b_i=p_i^*b$, where $p_i$, $i=1,2$, are the canonical projections.

In physical terms, as will be shown below, the $b_i$'s can be represented by SSH chains in 1D, and the product class
\begin{align}
b_1 b_2 \in \widetilde{K}^0(\BZ^2)\subset K^0(\BZ^2)\subset K^*(\BZ^2)
\end{align}
can be represented by a Chern insulator such as the anomalous Haldane insulator or a massive Dirac model.

The SSH chain in the non trivial phase yields a generator of $\widetilde{K}^{-1}(\BZ^1)$. To see this note that, in momentum space, the SSH Hamiltonian is specified by the continuous family
\begin{align}
H(k)=\left[\begin{array}{cc}
0 & v+ w e^{-ik}\\
v+w e^{ik} & 0
\end{array}\right],
\end{align}
where $k\in \BZ^1\cong S^1$, and $v,w$ are, respectively, hopping amplitudes. Observe that $H$ anti-commutes with 
\begin{align}
\Gamma=\left[\begin{array}{cc}
1 & 0\\
0 & -1
\end{array}\right].
\end{align}
Whenever $|w|>|v|$, we can continuously deform this family to
\begin{align}
\label{eq:SSH chain}
H(k)=\left[\begin{array}{cc}
0 & e^{-ik}\\
e^{ik} & 0
\end{array}\right],
\end{align}
and this defines an element $\text{SSH}=\{U(k)=e^{ik}\}_{k\in S^1}$. Actually, this is a generator of the first homotopy group of $\mbox{U}(1)$ (winding number $1$) and of the direct limit $\lim\limits_{\longrightarrow}\mbox{U}(N)$. As a consequence $\text{SSH}$ can be seen as a generator of $\pi_1 \mbox{U}\cong [S^1,\mbox{U}] \cong K^{-1}(S^1)\cong \widetilde{K}^{-1}(S^1)$.

To apply the definition of the product described in the beginning of this section in Eq.~\eqref{eq: product in k-cohomology}, we use $\widetilde{K}^{-1}(S^1)\cong\widetilde{K}^{0}(SS^1)\cong \widetilde{K}^0(S^2)$. Explicitly, we use the construction of Sec.~\ref{subsubsec: K-1 and suspensions}, to write
\begin{align}
&H(t,k)=\left[\begin{array}{cc}
\cos(\pi t) & \sin(\pi t)e^{-ik}\\
\sin(\pi t)e^{ik} & -\cos(\pi t) 
\end{array}\right]\nonumber\\
&= \sin(\pi t)\cos(k) \sigma_x + \sin(\pi t)\sin(k) \sigma_y +\cos(\pi t)\sigma_z,
\end{align}
which yields the usual Dirac monopole Hamiltonian of Example~\ref{example:product of dirac monopoles}. This intermediate step, using the construction of Sec.~\ref{subsubsec: K-1 and suspensions}, is also well known in condensed-matter literature and it is usually referred to as \emph{dimensional reduction}~\cite{teo:kan:10}. As a consequence, the $\text{SSH}$ class corresponds to the Dirac monopole class or, equivalently, the Bott class $b=[\mathcal{L}]-[1]\in \widetilde{K}^0(S^2)$.

Now the sequence of spaces $S^1\vee S^1\hookrightarrow S^1\times S^1\to S^1\wedge S^1=S^2$ induces an isomorphism $\widetilde{K}^0(\BZ^2)\cong \widetilde{K}^0(S^2)$ (see Hatcher's book~\cite{hat:03}, for example), where the last map is the quotient. Meanwhile, the product class
\begin{align}
&p_1^* \text{SSH} \; p_2^*\text{SSH} \nonumber\\
&=p_1^* b \; p_2^*b \in \widetilde{K}^0\left( \left(S^1\wedge S^1\right) \wedge \left(S^1\wedge S^1\right)\right) =\widetilde{K}^0(S^4),
\end{align}
is the generator of $\widetilde{K}^0(S^4)\cong \widetilde{K}^0(S^2)$, by Bott periodicity. Observe, however, that in the definition of the product we have to swap two $S^1$'s, so actually what we get is minus the Bott class over $S^1\wedge S^1$ which produces the desired product class in $\widetilde{K}^0(\BZ^2)$.

Before describing the obtained product class, it will be useful to illustrate the result that $p_1^*\text{SSH}\; p_2^*\text{SSH}$ generates $\widetilde{K}^0(S^4)$ more explicitly.  Note that, in terms of stable equivalence classes of Hamiltonians, we are computing the product class as described by the continuous family
\begin{align}
\label{eq: external tensor product of two dirac monopoles}
H(x_1)*H(x_2) &= H(x_1)\otimes H(x_2)\oplus \left(-H(x_1)\right)\otimes\sigma_3 \nonumber\\
&\oplus \sigma_3\otimes\left(-H(x_2)\right),  
\end{align}
with $x_i=[(t_i,k_i)]\in S S^1\cong S^2$, $i=1,2$, and $H(x)$ is the Dirac monopole Hamiltonian of Eq.~\eqref{eq: Dirac monopole}. We can parametrize each copy of $S^2$ using the Cartesian coordinates $x_i=(x_i^1,x_i^2,x_i^3)\in S^2\subset \mathbb{R}^3$, $i=1,2$. Now from the discussion of Appendix~\ref{sec:2-fold bott periodicity}, it follows that we can alternatively describe the stable equivalence class of the above Hamiltonian in terms of a generalized Dirac monopole as in Eq.~\eqref{eq: generalized Dirac monopole}:
\begin{align}
\widetilde{H}(x_1,x_2)&=x_1^1\sigma_1\otimes H(x_2) +x_1^2\sigma_2\otimes I_2  +x_1^3\sigma_3\otimes I_2 \nonumber\\
& =\sum_{j=1}^{3}x_1^1x_2^j\sigma_1\otimes \sigma_j +x_1^2\sigma_2\otimes I_2  +x_1^3\sigma_3\otimes I_2.
\end{align}
Observe that the matrices $\gamma_{i}=\sigma_1\otimes \sigma_i$, $i=1,..,3$, together with $\gamma_4=\sigma_2\otimes I_2$ and $\gamma_5=\sigma_3\otimes I_2$ satisfy
\begin{align}
\gamma_i\gamma_j+\gamma_j\gamma_i=\delta_{ij}I_4, \text{ for } i,j=1,...,5. \nonumber
\end{align}
Observe that $\gamma_5=-\gamma_1\gamma_2\gamma_3\gamma_4$. Moreover the five-dimensional vector appearing in $\widetilde{H}(x_1,x_2)$,
\begin{align}
y=(x_1^1 x_2^1,x_1^1x_2^2, x_1^1x_2^3, x_1^2,x_1^3)\in\mathbb{R}^5,
\end{align}
satisfies, $|y|^2=1$, and thus parametrizes a sphere $S^4\cong (S^1\wedge S^1)\wedge (S^1\wedge S^1)$. We have reduced the Hamiltonian to the usual Dirac form:
\begin{align}
\widetilde{H}(y)=\sum_{i=1}^{5}y^i\gamma_i,\  y\in S^4.
\end{align}
To see that this is a generator for $\widetilde{K}^0(S^4)$, note that isomorphism class of a vector bundle over $S^4$ is equivalently described by the homotopy class of the transition function in the equator of the sphere, which, in turn, is, in the stable sense, captured by the second Chern number of the bundle. It is enough to show that this number is $\pm 1$. This calculation is performed in Appendix~\ref{sec:2nd chern number for the external tensor product of dirac monopoles}. Swapping two $S^1$ factors can be achieved, modulo homotopy, by taking a reflection in any of the coordinates. This simply inverts the sign of the Chern number which becomes $+1$. This concludes the alternative proof that $p_1^*\text{SSH} \; p_2^*\text{SSH}$ is a generator for $\widetilde{K}^0(S^4)$. Using the Bott periodicity isomorphism, we can ``divide'' by the Bott class, and we know that this class corresponds to, up to a minus sign, that of the Dirac monopole over $S^1\wedge S^1\cong S^2$ -- after all, it was explicitly the external tensor product of two Dirac monopoles associated with two independent two-spheres. Notice, however, that the circles now correspond the one-dimensional independent Brillouin zones. To get the corresponding class in $\BZ^2$, we still need to pullback by the quotient map $q: S^1\times S^1\cong \BZ^2\to S^2$, which will not affect the topological invariant, in this case the first Chern number which is equal to $+1$, as argued below. 

We now show that the resulting product class can be represented by any Chern insulator of topological charge $+1$. To see this, suppose that $X$ is a smooth compact connected manifold. Take
\begin{align}
\label{eq:even chern character}
\textnormal{Ch}:& \; \widetilde{K}^0(X)\to H^{\text{even}}(X;\mathbb{R})\nonumber\\
 &[E]-[\theta^n]\mapsto \textnormal{Ch}(E) -n,
\end{align}
where $\textnormal{Ch}(E)$ is the Chern character of the bundle. Explicitly, if we are given a connection, such as the Berry connection for a subbundle of a trivial bundle, we can write the even de Rham class represented by the closed differential form
\begin{align}
\tr\; e^{i\frac{F}{2\pi}}=\sum_{k=0}^{\infty} \left(\frac{i}{2\pi}\right)^k\frac{1}{k!}\tr\; F^{k},
\end{align} 
where $F$ is the curvature of the connection. Then $\textnormal{Ch}$ commutes with pullbacks and it is an Abelian group homomorphism. Moreover, for the case of $X=S^2$ or $X=T^2$ it reduces to giving the first Chern class:
\begin{align}
\textnormal{Ch}(E)-n=n+c_1(E)-n=c_1(E)\in H^2(X;\mathbb{Z}).
\end{align}
Since the first Chern class of $\mathcal{L}$ is a generator of cohomology of $S^2$, namely $\int_{S^2} c_1(\mathcal{L})=-1$, it follows that the first Chern number provides an isomorphism $\widetilde{K}^0(S^2)\cong \mathbb{Z}$. The isomorphism $\widetilde{K}(T^2)\cong \widetilde{K}(S^2)$ is given by taking the pullback by the quotient $q: \BZ^2=T^2\to S^1\wedge S^1$ which has degree one (it is a homeomorphism on the complement of $S^1\vee S^1$) and hence the Chern number of the resulting bundle $q^*\mathcal{L}$ is the same -- thus, the first Chern number provides an isomorphism $\widetilde{K}^0(T^2)\cong\mathbb{Z}$. Now instead of $q^*\mathcal{L}$ we can choose any bundle with Chern number $-1$. The massive Dirac model is represented by the family
\begin{align}
&H(k_1,k_2)\nonumber\\
&=\sin(k_1)\sigma_1 +\sin(k_2)\sigma_2 + (M-\cos(k_1)-\cos(k_2))\sigma_3,
\end{align}
for $M=1$ the occupied bundle $\text{Im}\;p\to \BZ^2$ has first Chern number $+1$, and thus gives a generator of $\widetilde{K}(T^2)$. Let us denote by $p_i^*\text{SSH}=\text{SSH}_i$, $i=1,2$, and $[\text{Im}\;p]-[1]=\text{CH}$. The final result is
\begin{align}
\text{SSH}_1\; \text{SSH}_2 =\text{CH},
\end{align}
which can be elegantly stated as the product of two SSH chains (1D topological phases) is a Chern insulator (2D topological phase).
More generally, we have the result that
\begin{align}
K^{*}(\BZ^d)=\Lambda(\text{SSH}_1,...,\text{SSH}_d),
\end{align}
with $SSH_i=p_i^{*}SSH$, with $p_i$ the $i$th canonical projection to $\BZ^1\cong S^1$, $i=1,..,d$. So, the complex gapped topological phases of free fermions in $d$-dimensional translation invariant models are generated by $\text{SSH}$ chains, one for each independent direction.
\end{example}
\subsection{Relation to the topological invariant description}
\label{subsec: relation to the topological invariant description}
We will finish this section by making contact with the topological invariant description of complex topological phases of free fermions. This will be done through the Chern character graded ring isomorphism, relating complex \emph{K}-theory with the de Rham cohomology, which we proceed to explain. In Eq.~\eqref{eq:even chern character}, the \emph{even} Chern character for compact smooth manifolds $X$, actually provides a group homomorphism to $H^{\text{even}}(X;\mathbb{Q})$, meaning that when integrated over closed submanifolds of $X$, it yields rational numbers (this is because the Chern classes have values in $H^{\text{even}}(X;\mathbb{Z})$). One can extend the map to $\widetilde{K}^{-1}(X)$ by assigning to each class $[U]$, with $U:X\to \mbox{U}$, the closed odd differential form
\begin{align}
&\text{Ch}(U)\nonumber\\
&=\sum_{k=0}^{\infty}(-1)^{k}\left(\frac{i}{2\pi}\right)^{k+1}\frac{k!}{(2k+1)!}\tr\;\left[\left(U^{-1}dU\right)^{2k+1}\right],
\end{align} 
whose de Rham class is known as the \emph{odd} Chern character. Observe that for the case of the SSH chain in Eq.~\eqref{eq:SSH chain}, this just yields
\begin{align}
\text{Ch}(e^{ik})=\frac{i}{2\pi} e^{-ik}de^{ik}=-\frac{1}{2\pi}dk,
\end{align}
whose integral over the Brillouin zone yields minus the winding number of the map $\BZ^1\ni k\mapsto e^{ik}\in S^1$, equal to $-1$.

One can show that the odd Chern character provides a group homomorphism $\text{Ch}:\widetilde{K}^{-1}(X)\to H^{\text{odd}}(X;\mathbb{Q})$. One can extend it to the unreduced \emph{K}-theory by applying the formulas obtained to the case $X_{+}=X\sqcup\text{pt}$ -- equivalently, since $\widetilde{K}^{-1}(X)\cong K^{-1}(X)$, it is enough to extend Eq.~\eqref{eq:even chern character} to $K^{0}(X)$ by $\text{Ch}([E]-[F])=\text{Ch}(E)-\text{Ch}(F)\in H^{\text{even}}(X;\mathbb{Q})$, for any $[E]-[F]\in K^{0}(X)$. In fact, one can show that, by tensoring with the rationals to avoid torsion, the group homomorphism $\text{Ch}:K^*(X)\otimes \mathbb{Q}\to H^{*}(M ;\mathbb{Q})$ is actually a \emph{graded ring isomorphism} (see Refs.~\cite{par:08,kar:08}), where the ring structure in cohomology is the cup product induced by the exterior product of differential forms and the $\mathbb{Z}_2$ grading refers to even- and odd-degree differential forms. 

Before applying the isomorphism to derive consequences, at the level of the familiar topological invariants, from our main results, it may be useful to relate the even and odd Chern characters. This relation is obtained by means of the suspension. In fact, we have $\widetilde{K}^{0}(SX)=\widetilde{K}^{-1}(X)$ and $\widetilde{K}^{-1}(SX)=\widetilde{K}^{0}(SSX)\cong \widetilde{K}^{0}(X)$ (by Bott periodicity). Moreover, if we write a decomposition of the suspension in terms of cones, $SX=C_{-}X\cup C_{+}X$, then the connecting homomorphism, which in this case is an isomorphism, $\delta: H^{k}(X)\cong H^{k+1}(SX)$, for $k>0$, of the associated Mayer-Vietoris sequence in cohomology will transform the odd Chern character into the even Chern character. In terms of differential forms, this connecting homomorphism can be understood as follows. Fixing $k>0$, by contractibility of the cones, 
\begin{align}
\label{eq: even chern suspension}
\frac{1}{k!}\tr\; \left(\frac{i F}{2\pi}\right)^{k}\Big|_{C_{\pm}X}=dQ^{2k-1}_{\pm}, 
\end{align}  
for some $(2k-1)$-forms $Q^{2k-1}_{+}$ and $Q^{2k-1}_{-}$ defined over $C_+X$ and $C_{-}X$, respectively. Over the intersection $C_{+}X\cap C_{-}X\cong X$, since the gauge fields will differ by a gauge transformation $U$ defined over $X$, a standard calculation, see, for example, Ref.~\cite{gau:gin:89}, yields, up to an exact form,
\begin{align}
\label{eq: odd chern suspension}
&Q^{2k-1}_{+}-Q^{2k-1}_{-}\nonumber\\
&=(-1)^{k-1}\left(\frac{i}{2\pi}\right)^{k}\frac{(k-1)!}{(2k-1)!}\tr\left[\left(U^{-1}dU\right)^{2k-1}\right].
\end{align}
The connecting homomorphism $\delta: H^{2k-1}(X)\to H^{2k}(SX)$ sends the de Rham class of right-hand side of the above equation, Eq.~\eqref{eq: odd chern suspension}, to the de Rham class of the left-hand side of Eq.~\eqref{eq: even chern suspension}.
Observe that for a line bundle over a two-sphere, the above formula yields the familiar result that the first Chern number is a winding number of a transition function defined over the equator circle. Indeed, let us consider the  case of the SSH chain and the identification $\widetilde{K}^{-1}(\BZ^1)=\widetilde{K}^{0}(S\BZ^1)=\widetilde{K}^{0}(S^2)$. The cones $C_{\pm}\BZ^1$ are identified with neighbourhoods of the north and south poles of the two-sphere $S\BZ^1\cong S^2$. The relation between the Berry gauge fields over the cones $C_{\pm}\BZ^1$, denoted $A_{\pm}$, at the overlap, $C_{-}\BZ^1\cap C_{+}\BZ^1\cong \BZ^1$, is
\begin{align}
A_{+}-A_{-} =\text{Ch}(e^{ik})=-\frac{1}{2\pi} dk,
\end{align}
yielding that the first Chern number is equal to $+1$, which is equal to the winding number of $e^{ik}$.

Finally, because $\text{Ch}:K^*(X)\otimes \mathbb{Q}\to H^{*}(M ;\mathbb{Q})$ is a graded ring isomorphism, we can read off the relation between the topological invariants in different dimensions. For example, from our result $\text{SSH}_1\text{SSH}_2=\text{CH}$, we can apply $\text{Ch}$ to both sides of the equation to obtain
\begin{align}
\text{Ch}\left(\text{SSH}_1\text{SSH}_2\right)=\text{Ch}\left(\text{SSH}_1\right)\smile \text{Ch}\left(\text{SSH}_2\right)=\text{Ch}(\text{CH}),
\end{align}
where $\smile$ is the cup product in cohomology. Since, at the level of differential forms,
\begin{align}
\text{Ch}\left(\text{SSH}_i\right)=-\frac{1}{2\pi} dk_i, \ i=1,2,
\end{align}
integration over $\BZ^2=\BZ^1\times \BZ^1$ yields
\begin{align}
&\int_{\BZ^1\times \BZ^1}\left(-\frac{i}{2\pi} dk_1\right)\smile \left(-\frac{i}{2\pi} dk_2\right)\nonumber\\
&=\int_{\BZ^1\times \BZ^1}\left(-\frac{i}{2\pi} dk_1\right)\wedge \left(-\frac{i}{2\pi} dk_2\right)\nonumber\\
&=\int_{\BZ^1}\left(-\frac{i}{2\pi} dk_1\right)\int_{\BZ^1}\left(-\frac{i}{2\pi} dk_2\right)\nonumber\\
&= 1= \int_{\BZ^1\times \BZ^1}\text{Ch}(\text{CH}),
\end{align}
which gives the topological charge one, as described previously. Hence, the relation between the topological invariants of $\text{SSH}_1$, $\text{SSH}_2$, and $\text{CH}$ is that the invariant for the latter is the product of the invariants for the two former ones. Thus, we conclude that our main result provides not only a conceptual advance in the understanding of the product structure, but also provides the necessary means, through the Chern character ring isomorphism, to completely determine and relate the usual topological invariants in different dimensions.

\section{Conclusions}
\label{sec:conclusions}
We have provided an in-depth review of \emph{K}-theory and $\mod 2$-Bott periodicity in the context of complex gapped phases of free fermions while emphasizing on a product structure which was previously neglected in literature because its physical interpretation was not clear. In Sec.~\ref{sec: Kunneth formula and SSH times SSH equals Chern insulator}, we introduced the \emph{K}-cohomology group together with the associated ring structure, allowing us to derive, using K\"unneth's theorem, the result that the SSH chains, one for each independent momentum direction, generate the \emph{K}-cohomology groups of the Brillouin zone in $d$ dimensions and that the product of two SSH chains is a Chern insulator. These results relate the associated topological phases and their topological invariants in all spatial dimensions in a unified way.

One could wonder if similar results hold for the case of topological phases of gapped free fermions as described by real \emph{K}-theory. Unfortunately, in that case, the external tensor product is not always injective and this implies that the K\"{u}nneth formula does not hold for the real \emph{K}-theory, see Atiyah's discussion in Ref.~\cite{ati:62}. Nevertheless, multiplicative structures are fundamental to understand the periodic table of topological insulators and superconductors and the associated anomalous surface states, i.e., complex and real \emph{K}-theory, and, more generally, of understanding crystalline topological insulators and superconductors~\cite{cor:car:20}.

We hope these results shed new light on the physics of this product structure -- previously thought of as just mathematical in nature, devoid of physical character -- and motivates further results in the classification of topological phases within the same spirit.

\section*{Acknowledgements}
\label{sec: Acknowledgements}

B.M. acknowledges stimulating discussions with J. P. Nunes. B.M. is thankful for the support from SQIG -- Security and Quantum Information Group, under the Funda\c{c}\~ao para a Ci\^{e}ncia e a Tecnologia (FCT) project UIDB/50008/2020, and European funds, namely, H2020 project SPARTA. B.M. acknowledges projects QuantMining POCI-01-0145-FEDER-031826, PREDICT PTDC/CCI-CIF/29877/2017 and an internal IT project, QBigData PEst-OE/EEI/LA0008/2013, funded by FCT. 

\appendix
\section{The twofold Bott periodicity of complex \emph{K}-theory}
\label{sec:2-fold bott periodicity}
In the following, we will see the twofold Bott periodicity of complex \emph{K}-theory from a different perspective using the suspension construction, and show how the Dirac monopole or, equivalently, the tautological bundle over the sphere $\mathcal{L}\to S^2$ plays an important role in it. In particular, what we will show is that $\widetilde{K}^{-2}(X)$ which, under the above definition, is naturally identified with $\widetilde{K}^{0}(X)$ is isomorphic to $\widetilde{K}^{0}(X\wedge S^2)=\widetilde{K}^{0}(\Sigma^2X)$, which is the usual definition of $\widetilde{K}^{-2}(X)$. 

For the purpose of the discussion, it is convenient to define $K^{-2}(X)$ in terms of the families of matrices with two Clifford symmetries. So we will take the families for the form of Eq.~\eqref{Eq:Hamiltonian 2-Clifford symmetric}. Consider then the inclusion
\begin{align}
\left[\begin{array}{cc}
0_{N} &  h\\
h & 0_{N}
\end{array}\right]\mapsto \left[\begin{array}{cc}
0_{N+1} &  h\oplus 1\\
h\oplus 1 & 0_{N+1}
\end{array}\right], \text{ with } N>0,
\end{align}
take the direct limit, with respect to these inclusions, and quotient by the equivalence relation of homotopy preserving the Clifford symmetries. The resulting set has the structure of an Abelian monoid under the operation
\begin{align}
&\left(\left[\begin{array}{cc}
0_{N_1} &  h_1\\
h_1 & 0_{N_1}
\end{array}\right],\left[\begin{array}{cc}
0_{N_2} &  h_2\\
h_2 & 0_{N_2}
\end{array}\right]\right)\mapsto \left[\begin{array}{cc}
0_{N_1+N_2} &  h_1\oplus h_2\\
h_1\oplus h_2 & 0_{N_1+N_2}
\end{array}\right],\nonumber\\
& \text{with } N_1,N_2>0.
\end{align}
Performing the Grothendieck group completion yields an Abelian group which we call $K^{-2}(X)$. The elements of $K^{-2}(X)$ can be thought of as differences $[\{H_1(x)\}_{x\in X}]-[\{H_2(x)\}_{x\in X}]$ and by the same argument as in $K^0(X)$, we can bring the second family to a trivial form which, in this case, is given by
\begin{align}
H_{\text{trivial},N}=\left[\begin{array}{cc}
0_{2N} &  I_N\oplus (-I_N)\\
I_N\oplus(-I_N) & 0_{2N}
\end{array}\right],
\end{align}
for some $N\in\mathbb{N}$. Again, one can consider the kernel of the induced map $i^*:K^{-2}(X)\to K^{-2}(\{x_0\})$ and this provides the reduced \emph{K}-theory group $\widetilde{K}^{-2}(X)$, whose elements can be thought of differences $[H]-[H_{\text{trivial},N}]$, where $N$ is equal to the rank of eigenbundle associated with the $-1$ eigenvalue of $h$, $\mbox{Im}\;p\to X$. 

Suppose we have a family $\{H(x)\}_{x\in X}$ satisfying the two required Clifford symmetries and, thus, being defined by $\{h(x)\}_{x\in X}$ as in Eq.~\eqref{Eq:Hamiltonian 2-Clifford symmetric}. Then we can define a family with a single Clifford symmetry over the suspension $SX$ by
\begin{align}
&H(t,x)=\cos(\pi t)\Gamma_2 +\sin(\pi t) H(x) \nonumber\\
&=\left[\begin{array}{cc}
0_N & -i \cos(\pi t)I_N \!+\! \sin(\pi t)h(x)\\
i \cos(\pi t)I_N \!+\! \sin(\pi t)h(x)    & 0_N
\end{array}\right],\nonumber\\
&\text{ for } t\in [0,1],\ x\in X.
\end{align}
Observe that
\begin{align}
U(t,x)=i \cos(\pi t)I_N +\sin(\pi t)h(x)
\end{align}
satisfies $U(0,x)=i I_N$, $U(1,x)=-i I_N$, for all $x\in X$,
\begin{align}
&U^\dagger(t,x) U(t,x)=\left(\cos^2(\pi t)+\sin^2(\pi t)\right) I_N =I_N,\nonumber\\
& \text{ for all } (t,x)\in [0,1]\times X.
\end{align}
Hence, $\{U(t,x)\}_{[(t,x)]\in SX}$ is a well-defined continuous family of unitary $N\times N$ matrices over $SX$ and  $\{H(t,x)\}_{[(t,x)]\in SX}$ is a well-defined continuous family of Hamiltonians with a single Clifford symmetry $\Gamma_1$. Observe that under the inclusions above, we have $h(x)\mapsto h(x)\oplus 1$ and $U(t,x)\mapsto U(t,x)\oplus 1$. Thus, we have a well-defined map
\begin{align}
\widetilde{K}^{-2}(X)\ni [H]-[H_{\text{triv},n}]\mapsto [U]\in K^{-1}(SX),
\end{align}
or, equivalently,
\begin{align}
\widetilde{K}^{0}(X)\ni [\mbox{Im}\;p]-[\theta^n]\mapsto [U]\in K^{-1}(SX), 
\end{align}
where $n$ is the rank of the eigenbundle $\mbox{Im}\; p$. The map is a group homomorphism. Note the similarity of this construction with that in Sec.~\ref{subsubsec: K-1 and suspensions}, but here with $\Gamma_2$ instead of $\Gamma$. Again, it is useful to extend $U(t,x)$ to a loop of unitaries, or equivalently, to a loop of Hamiltonians with one Clifford symmetry. To do this, we just observe that $U(1,x)=-i I_N$, independently of $x$. We can then extend by declaring that, for $1\leq t\leq 2$,
\begin{align}
U(t,x)=i\cos(\pi t)I_N +\sin(\pi t)(-I_{n}\oplus I_{N-n}),
\end{align}
where $n$ equals the number of negative eigenvalues of $h$. The construction described above is, up to multiplication by $i$ and when we take $N\mapsto 2N$ and $n\mapsto N$ (which can be done under stable isomorphism), precisely applying the map $f:G_N=\mbox{U}(2N)/\mbox{U}(N)\times \mbox{U}(N)\to \Omega U(2N)$ described by Bott in Ref.~\cite{bot:59}. Indeed, if we write $h(x)=q(x)-p(x)$, where $p(x)=\Theta(-h(x))$ is the associated orthogonal projector, and $q(x)=I_{2N}-p(x)$, we have that
\begin{align}
U(t,x)=\begin{cases}
ie^{i\pi t} p(x) + ie^{-i\pi t} q(x) ,\ t\in[0,1],\\
ie^{i\pi t} p_0 + ie^{-i\pi t} q_0  ,\ t\in [1,2],
\end{cases}
\end{align} 
where 
\begin{align}
p_0=I_{N}\oplus 0_N \text{ and } q_0= 0_N\oplus I_N.
\end{align}
Observe then, with the two maps $\lambda:\mbox{U}(2N)\to \Omega G_{2N}$ and $f: G_{N}\to \Omega U(2N)$, we have a map, as defined by Bott~\cite{bot:59}, $\gamma= \Omega \lambda \circ f: G_{N}\to \Omega^2 G_{2N}$ and its adjoint $\gamma^*: G_N\wedge S^2 \to G_{2N}$, given by
\begin{align}
\gamma(h) (t_1,t_2)=\gamma^*([(h,[(t_1,t_2)])])= \lambda (f(t_1))(t_2).
\end{align}
In Bott's original work~\cite{bot:57, bot:59}, he showed, making use of Morse theory, that in the $N\to \infty$ limit, $\gamma$ is an homotopy equivalence. The direct limit of $G_{N}$ is, up to homotopy, $\mbox{BU}$, the classifying space for the unitary group. The classifying space satisfies that the homotopy classes of maps $[X,\mbox{BU}]=\mathcal{E}\mathcal{U}(X)$, since  homotopy classes of maps from $X$ to the spaces of projectors modulo stable equivalence is in bijection with stable equivalence classes of vector bundles over $X$. As a consequence of Bott's work, it follows that $[X,\mbox{BU}]\cong [X, \Omega^2 \mbox{BU}]$. Because, by duality, $[X,\Omega^k Y]\cong [X\wedge S^k, Y]$, for natural $k$, we have $[X,\mbox{BU}]\cong [X\wedge S^2,\mbox{BU}]$, i.e., $\widetilde{K}^0(X)\cong \widetilde{K}^0(X\wedge S^2)$, through $\gamma$. Bott showed in Ref.~\cite{bot:59} also that $\gamma$ is homotopy equivalent to the map induced by external tensor product by the Bott class $b=[\mathcal{L}]-[\theta^1]\in \widetilde{K}(S^2)$.
 
Note that $\gamma$ defines the Hamiltonian
\begin{align}
&H(t_1,t_2,x)=\cos(\pi t_1)\Gamma_1 +\sin(\pi t_1)H(t_2,x),\nonumber\\
&\text{ where } t_1\in [0,1], \text{ and }[(t,x)]\in SX.
\end{align}
The formula above defines a family over $S^2X=S(SX)$, the \emph{double suspension of} $X$ with \emph{no Clifford symmetries}. We will denote elements of $S^2X$ by their equivalence classes $[(t_1,t_2,x)]$, with $(t_1,t_2,x)\in [0,1]^2\times X$. In \emph{K}-theory terms, this procedure corresponds to applying the isomorphism $K^{-1}(Y)\cong \widetilde{K}^0(SX)$ constructed in Sec.~\ref{subsubsec: K-1 and suspensions}, with $Y=SX$. In other words, we have built a map
\begin{align}
\widetilde{K}^{-2}(X)\ni [H(x)]-[H_{\text{triv},N}]\mapsto [P]-[I_N]\in \widetilde{K}^0(S^2X),
\end{align} 
where $2N$ is the size of the Hermitian matrix $h$ appearing in $H$ and $P([(t_1,t_2,x)])=\Theta(-H(t_1,t_2,x))$ for all $[(t_1,t_2,x)]\in S^2X$. 

We have cones $C_{-}(SX)$ and $C_{+}(SX)$, restricted to which the associated bundle $\mbox{Im}\;P$ trivializes. If we introduce the $2N\times 2N$ matrices
\begin{align}
Z(t_1,t_2,x)&=\tan\left(\frac{\pi t_1}{2}\right) U(t_2,x)\nonumber\\
&=\tan\left(\frac{\pi t_1}{2}\right)\left(i \cos(\pi t_2)I_{2N} +\sin(\pi t_2)h(x)\right),\nonumber\\
&\text{with } [(t_1,t_2,x)]\in C_{-}(SX),
\end{align}
and 
\begin{align}
W(t_1,t_2,x)&=\cot\left(\frac{\pi t_1}{2}\right) U^\dagger(t_2,x)\nonumber\\
&=\cot\left(\frac{\pi t_1}{2}\right)\left(-i \cos(\pi t_2)I_N +\sin(\pi t_2)h(x)\right),\nonumber\\
&\text{with } [(t_1,t_2,x)]\in C_{+}(SX),
\end{align}
the eigenvectors with eigenvalue $+1$ are described by the unitary $2N$ frames
\begin{align}
&v_{-}(t_1,t_2,x)\nonumber\\
&=\left[\begin{array}{cc}
I_{2N}\\
Z(t_1,t_2,x)
\end{array}
\right] (I_{2N}+Z^\dagger(t_1,t_2,x)Z(t_1,t_2,x))^{-1/2},\nonumber\\
&\text{for } [(t_1,t_2,x)]\in C_{-}(SX),
\end{align} 
and
\begin{align}
&v_{+}(t_1,t_2,x)\nonumber\\
&=\left[\begin{array}{cc}
W(t_1,t_2,x)\\
I_{2N}
\end{array}
\right] (I_{2N}+W^\dagger(t_1,t_2,x)W(t_1,t_2,x))^{-1/2},\nonumber\\
&\text{for } [(t_1,t_2,x)]\in C_{+}(SX).
\end{align}
The relation over $C_{-}(SX)\cap C_{+}(SX)=\{1/2\}\times SX\cong SX$ is given by
\begin{align}
v_{-}(1/2,t_2,x)=v_{+}(1/2,t_2,x) U(t_2,x),
\end{align}
and this follows since
\begin{align}
&Z(1/2,t_2,x)=U(t_2,x)=W^{-1}(1/2,t_2,x),\nonumber\\
&\text{for } [(t_2,x)]\in SX.
\end{align}
This relation does in fact hold everywhere except for $t_1=0$ and $t_1=1$, i.e., whenever both matrices are simultaneously defined. Denote by $p(x)=\Theta(-h(x))$ and by $q(x)=I_{2N}-p(x)$, the orthogonal complement. Observe that
\begin{align}
&Z(t_1,t_2,x)=\tan\left(\frac{\pi t_1}{2}\right)U(t_2,x)\nonumber\\
&=\tan\left(\frac{\pi t_1}{2}\right)\left(i\cos(\pi t_2)I_{2N}+\sin(\pi t_2)h(x)\right)\nonumber\\
&=\tan\left(\frac{\pi t_1}{2}\right)\big(i\cos(\pi t_2)(p(x)+q(x))\nonumber\\
&+\sin(\pi t_2)(q(x)-p(x))\big)\nonumber\\
&=i\tan\left(\frac{\pi t_1}{2}\right)e^{i\pi t_2}p(x)+i\tan\left(\frac{\pi t_1}{2}\right)e^{-i\pi t_2}q(x),
\end{align}
and
\begin{align}
&W(t_1,t_2,x)=\cot\left(\frac{\pi t_1}{2}\right)U^\dagger(t_2,x)\nonumber\\
&=\cot\left(\frac{\pi t_1}{2}\right)\left(-i\cos(\pi t_2)I_{2N}+\sin(\pi t_2)h(x)\right)\nonumber\\
&=\cot\left(\frac{\pi t_1}{2}\right)\big(-i\cos(\pi t_2)(p(x)+q(x))\nonumber\\
&+\sin(\pi t_2)(q(x)-p(x))\big)\nonumber\\
&=-i\cot\left(\frac{\pi t_1}{2}\right)e^{-i\pi t_2}p(x)-i\cot\left(\frac{\pi t_1}{2}\right)e^{i\pi t_2}q(x).
\end{align}
Now define
\begin{align}
z(t_1,t_2)=i\tan\left(\frac{\pi t_1}{2}\right)e^{i\pi t_2}\in\mathbb{C}.
\end{align}
Similarly, define
\begin{align}
w(t_1,t_2)=-i\cot\left(\frac{\pi t_1}{2}\right) e^{-i\pi t_2}\in\mathbb{C}.
\end{align}
Observe that if $t_1\neq \{0,1\}$, we can write $z=1/w$.
Then, we see that,
\begin{align}
&v_{-}(t_1,t_2,x)\nonumber\\
&=\left[\begin{array}{cc}
I_{2N}\\
Z(t_1,t_2,x)
\end{array}\right](I_{2N}+Z^\dagger(t_1,t_2,x)Z(t_1,t_2,x))^{-1/2} \nonumber\\
&=\frac{1}{\sqrt{1+|z(t_1,t_2)|^2}}\left[\begin{array}{cc}
q(x)\\
z(t_1,t_2)q(x)
\end{array}\right]\nonumber\\
&+\frac{1}{\sqrt{1+|z(t_1,t_2)|^2}}\left[\begin{array}{cc}
p(x)\\
-\bar{z}(t_1,t_2)p(x)
\end{array}\right]\nonumber\\
&=\frac{1}{\sqrt{1+|z(t_1,t_2)|^2}}\left[\begin{array}{cc}
1\\
z(t_1,t_2)
\end{array}\right]\otimes q(x)\nonumber\\
&+\frac{1}{\sqrt{1+|z(t_1,t_2)|^2}}\left[\begin{array}{cc}
1\\
-\bar{z}(t_1,t_2)
\end{array}\right]\otimes p(x),
\end{align}
and, similarly,
\begin{align}
&v_{+}(t_1,t_2,x)\nonumber\\
&=\left[\begin{array}{cc}
W(t_1,t_2,x)\\
I_{2N}
\end{array}\right](I_{2N}+W^\dagger(t_1,t_2,x)W(t_1,t_2,x))^{-1/2}\nonumber\\
&=\frac{1}{\sqrt{1+|w(t_1,t_2)|^2}}\left[\begin{array}{cc}
w(t_1,t_2)\\
1
\end{array}\right]\otimes q(x)\nonumber\\
&+\frac{1}{\sqrt{1+|z(t_1,t_2)|^2}}\left[\begin{array}{cc}
-\bar{w}(t_1,t_2)\\
1
\end{array}\right]\otimes p(x).
\end{align}

The structure of the two previous equations provides us a splitting of a bundle of rank $2N$, $E\to S^2\times X$ into two subbundles of rank $N$,
\begin{align}
E =E_{1}\oplus E_{2},
\end{align}
provided we allow $z$ and $w$ to range over the complex numbers and preserve the gluing condition $z=1/w$, whenever $z,w\neq 0$. We proceed to describe $E_1$ and $E_{2}$. The Riemann sphere $\mathbb{C}\cup\{\infty\}$ has the usual stereographic projection coordinate
\begin{align}
z=\tan\left(\frac{\theta}{2}\right)e^{i\phi},
\end{align} 
where $\theta\in[0,\pi]$ and $\phi\in[0,2\pi)$. By analogy with the above, we write
\begin{align}
&Z(z,x)=zq(x) -\bar{z} p(x)=\frac{z}{2}(I_{2N}+h(x)) -\frac{\bar{z}}{2}(I_{2N}-h(x))\nonumber\\
&=\frac{z-\bar{z}}{2}I_{2N} +\frac{z+\bar{z}}{2}h(x)\nonumber\\
&=\tan\left(\frac{\theta}{2}\right)\left(i\sin(\phi) I_{2N} +\cos(\phi) h(x)\right),
\end{align}
and, similarly,
\begin{align}
W(w,x)=wq(x)-\bar{w}p(x).
\end{align}
From which it is clear that in $\mathbb{C}-\{0\}\subset \mathbb{C}\cup\{\infty\}$,
\begin{align}
Z(z,x) W(w=1/z,x)=p(x) +q(x)=I_N.
\end{align}
The bundle $E\to S^2\times X$ is defined through the trivializations
\begin{align}
v_{-}(z,x)=\frac{1}{\sqrt{1+|z|^2}}\left[\begin{array}{cc}
I_{2N}\\
Z(z,x)
\end{array}\right] \text{ for } (z,x)\in \mathbb{C}\times X,
\end{align}
i.e., away from the north pole, and
\begin{align}
v_{+}(w,x)=\frac{1}{\sqrt{1+|w|^2}}\left[\begin{array}{cc}
W(w,x)\\
I_{2N}
\end{array}\right] \text{ for } (w,x)\in \mathbb{C}\times X,
\end{align}
i.e., away from the south pole. In the overlap, i.e., for $z\in \mathbb{C}-\{0\}$, the transition function is 
\begin{align}
v_{-}(z,x)=v_{+}(1/z,x) Z(z,x)(Z^\dagger(z) Z(x))^{-1/2}.
\end{align}
Now the bundles $E_{1}$ and $E_{2}$ come from the decompositions
\begin{align}
&v_{-}(z,x)\nonumber\\
&=\frac{1}{\sqrt{1+|z|^2}}\left[\begin{array}{cc}
1\\
z
\end{array}\right]\otimes q(x)+\frac{1}{\sqrt{1+|z|^2}}\left[\begin{array}{cc}
1\\
-\bar{z}\end{array}\right]\otimes p(x)
\end{align}
and
\begin{align}
&v_{+}(w,x)\nonumber\\
&=\frac{1}{\sqrt{1+|w|^2}}\left[\begin{array}{cc}
w\\
1
\end{array}\right]\otimes q(x)+\frac{1}{\sqrt{1+|w|^2}}\left[\begin{array}{cc}
-\bar{w}\\
1
\end{array}\right]\otimes p(x),
\end{align}
which, therefore, hold globally. The bundle $E_{1}$ corresponds to the first summand and the bundle $E_{2}$ to the second. 

Perhaps it might be useful to recall a few facts about the Dirac monopole bundle, see also Example~\ref{example:product of dirac monopoles}. In the language described in this paper, it corresponds to the family over $S^2\subset \mathbb{R}^3$ defined by
\begin{align}
H(x)= \vec{x}\cdot \vec{\sigma}=\left[\begin{array}{cc}
x^3 & x^1-ix^2\\
x^1+ix^3 & -x^3
\end{array}\right].
\end{align}
Using stereographic projection with respect to the south pole of $S^2$, we get the complex coordinate $z=(x^1+ix^2)/(1+x^3)$ and the $+1$ eigenvalue bundle can be described by
\begin{align}
v_{-}(z)=\frac{1}{\sqrt{1+|z|^2}}\left[\begin{array}{cc}
1\\
z
\end{array}\right] \text{ for } z\in\mathbb{C}.
\end{align}
This description misses the south pole of the sphere $z=\infty$, which is captured by
\begin{align}
v_{+}(w)=\frac{1}{\sqrt{1+|w|^2}}\left[\begin{array}{cc}
w\\
1
\end{array}\right] \text{ for } w\in\mathbb{C},
\end{align}
where $w$ is the stereographic projection with respect to the north pole. When the two descriptions are available, i.e., in $S^2-\{N,S\}\cong \mathbb{C}-\{0\}$, the relation between them is given by
\begin{align}
v_{-}(z)=v_{+}(1/z)\frac{z}{|z|},
\end{align}
hence the winding number $1$ transition map at the equator of the sphere given by $g_{+-}:\phi\mapsto e^{i\phi}\in \mbox{U}(1)$. This bundle is also known as the tautological bundle over the complex projective line $\mathbb{C}P^1\cong\mathbb{C}\cup\{\infty\}$, the space of all one-dimensional subspaces of $\mathbb{C}^2$, because the fiber over each point is precisely the point itself which is a one-dimensional subspace of $\mathbb{C}^2$. The dual bundle, $\overline{\mathcal{L}}$, has a transition map which is the complex conjugate of this one. We then conclude that
\begin{align}
E_{1}=p_1^{*}\mathcal{L} \otimes p_2^{*}\mbox{Im}\; q \text{ and } E_{2}=p_1^{*}\overline{\mathcal{L}}\otimes p_2^{*}\mbox{Im}\; p,
\end{align}
where $p_1:S^2\times X\to S^2$ and $p_2:S^2\times X\to X$ are the natural projections. I.e., from the bundle $\mbox{Im}\;p\to X$, we have constructed, through $\gamma$, a bundle
\begin{align}
E=p_1^{*}\mathcal{L} \otimes p_2^{*}\mbox{Im}\;q \oplus p_1^{*}\overline{\mathcal{L}}\otimes p_2^{*}\mbox{Im}\;p\to S^2\times X.
\end{align}
In terms, of Hamiltonians, $E$ can be described as a \emph{generalized Dirac monopole}. Namely, $E$ is the positive eigenvalue bundle of
\begin{align}
\label{eq: generalized Dirac monopole}
&H(x^1,x^2,x^3,x)\nonumber\\
&=x^1\sigma_1\otimes h(x)+x^2\sigma_2\otimes I_{2N} +x^3\sigma_3\otimes I_{2N}
\end{align}
where $(x^1,x^2,x^3)\in S^2\subset\mathbb{R}^3$ and $x\in X$.

Let us see that $E$ is isomorphic to the bundle induced by the map $\gamma^*:G_{N}\wedge S^2\to G_{2N}$. The latter bundle is built as follows. Over the Grassmannian $G_{N}=\mbox{U}(2N)/\left(\mbox{U}(N)\times \mbox{U}(N)\right)$, we have the tautological bundle whose fiber at an $N$-plane is the plane itself. Let us call this bundle $E^{2N}_{N}$. Then, the Hamiltonian $h$ defines a map $\hat{h}: X\to G_N$ by $x\mapsto \mbox{Im}\;q$, so that $f=\hat{h}\wedge \text{id}_{S^2}: X\wedge S^2\to G_{N}\wedge S^2$ (where $\text{id}_{S^2}$ is the identity over the $S^2$ factor) provides a map $\gamma^*\circ f: X\wedge S^2\to G_{2N}$. Then 
\begin{align}
F=\left(\gamma^*\circ f\right)^*E^{2N}_{N} \cong f^*\left( (\gamma^*)^*E^{2N}_{N}\right)
\end{align}  
is the bundle induced by $\gamma^*$. The bundle $F$ can be described as follows, see Ref.~\cite{bot:59} for more details. Take $S^2=D_{+}\cup D_{-}\cong\mathbb{C}P^1$ as the union of two disks whose intersection corresponds to the lines in $\mathbb{C}P^1$ which are invariant under complex conjugation. Then 
\begin{align}
F|_{D_{+}\times X}&=\left(p_1^{*}\mathcal{L} \otimes p_2^{*}\mbox{Im}\;q \oplus p_1^{*}\overline{\mathcal{L}}\otimes p_2^{*}\mbox{Im}\;p\right)\Big|_{D_{+}\times X },\nonumber\\
F|_{D_{-}\times X}&=\left(p_1^{*}\mathcal{L} \otimes p_2^{*}\theta^N \oplus p_1^{*}\overline{\mathcal{L}}\otimes p_2^{*}\theta^N\right)\Big|_{D_{-}\times X}
\end{align}
Observe that over $D_{\pm}$ the bundles $\mathcal{L}$ and $\overline{\mathcal{L}}$ are trivializable, yielding
\begin{align}
E|_{D_{-}\times X}&=\left(p_1^{*}\mathcal{L} \otimes p_2^{*}\mbox{Im}\;q \oplus p_1^{*}\overline{\mathcal{L}}\otimes p_2^{*}\mbox{Im}\;p\right)\Big|_{D_{-}\times X}\nonumber\\
&\cong \left(p_1^{*}\theta^1 \otimes p_2^{*}\mbox{Im}\;q \oplus p_1^{*}\theta^1\otimes p_2^{*}\mbox{Im}\;p\right)\Big|_{D_{-}\times X}\nonumber\\
&\cong\left(p_1^{*}\theta^1 \otimes \left(p_2^{*}\mbox{Im}\;q \oplus  p_2^{*}\mbox{Im}\;p\right)\right)\Big|_{D_{-}\times X}\nonumber\\
&\cong\left(p_1^{*}\theta^1 \otimes p_2^{*}\theta^N \oplus p_1^{*}\theta^1\otimes p_2^{*}\theta^N\right)\Big|_{D_{-}\times X}\nonumber\\
&\cong\left(p_1^{*}\theta^2 \otimes p_2^{*}\theta^N \right)\Big|_{D_{-}\times X}\nonumber\\
&\cong\left(p_1^{*}\left(\mathcal{L}\oplus \mathcal{L}^{\perp}\right) \otimes p_2^{*}\theta^N\right)\Big|_{D_{-}\times X}\nonumber\\
&\cong\left(p_1^{*}\left(\mathcal{L}\oplus \overline{\mathcal{L}}\right) \otimes p_2^{*}\theta^N\right)\Big|_{D_{-}\times X}\nonumber\\
&\cong\left(p_1^{*}\mathcal{L} \otimes p_2^{*}\theta^N \oplus p_1^{*}\overline{\mathcal{L}} \otimes p_2^{*}\theta^N \right)\Big|_{D_{-}\times X}\nonumber\\
&=F|_{D_{-}\times X},
\end{align}
where we used $\mathcal{L}^{\perp}\cong \overline{\mathcal{L}}$, because they share the same transition function. Moreover, at $\partial D_{-}=\partial D_{-}$ corresponding to the lines invariant under complex conjugation, we have that $\mathcal{L}|_{\partial D_{-}}=\bar{\mathcal{L}}|_{\partial D_{-}}$, and hence
\begin{align}
E|_{\partial D_{-}\times X} &=F|_{\partial D_{-}\times X}.
\end{align}
We conclude that the isomorphism over $D_{-}\times X$ extends to the whole of $S^2\times X$ by declaring that it is the identity over $D_{+}\times X$ -- hence the bundles $E$ and $F$ are isomorphic. Now we can show that this bundle is stably isomorphic to
\begin{align}
&p_1^*\mathcal{L}\otimes p_2^*\mbox{Im}\;q\oplus \theta^1\otimes p_2^*\mbox{Im}\; p\oplus p_1^*\mathcal{L}^{\perp}\otimes \theta^N\nonumber\\
&=p_1^{*}\mathcal{L} * p_2^*\mbox{Im}\;q,
\end{align}
representing the external tensor product of the Bott class $[\mathcal{L}]_s$ with $[\mbox{Im}\;q]_s$, since $\left(\mbox{Im}\;q\right)^{\perp}=\mbox{Im}\;p$. To see this, take $E$ and consider the stably isomorphic bundle
\begin{align}
&E\oplus \theta^{2}\otimes\theta^{N}\nonumber\\
&=p_1^{*}\mathcal{L} \otimes p_2^{*}\mbox{Im}\;q \oplus p_1^{*}\overline{\mathcal{L}}\otimes p_2^{*}\mbox{Im}\;p\oplus p_1^*\theta^{2}\otimes p_2^*\theta^{N}.
\end{align}
Now $\theta^2\cong \mathcal{L}\oplus \mathcal{L}^{\perp}$, so,
\begin{align}
\label{eq: E stable iso to part 1}
&p_1^{*}\mathcal{L} \otimes p_2^{*}\mbox{Im}\;q \oplus p_1^{*}\overline{\mathcal{L}}\otimes p_2^{*}\mbox{Im}\;p\oplus \theta^{2}\otimes\theta^{N}\nonumber\\
&\cong p_1^{*}\mathcal{L} \otimes p_2^{*}\mbox{Im}\;q \nonumber \\
&\oplus p_1^{*}\overline{\mathcal{L}}\otimes p_2^{*}\mbox{Im}\;p\oplus p_1^*\mathcal{L}\otimes p_2^*\theta^N\oplus p_1^*\mathcal{L}^{\perp}\otimes p_2^*\theta^N.
\end{align}
To proceed, we show that 
\begin{align}
\label{eq:a useful isomorphism}
p_1^{*}\overline{\mathcal{L}}\otimes p_2^{*}\mbox{Im}\;p\oplus p_1^*\mathcal{L}\otimes p_2^*\theta^N\cong p_2^{*}\left(\mbox{Im}\;p\oplus \theta^N\right).
\end{align}
We can form an open cover $\{U_{\alpha}\}_{\alpha=1}^{M}$ of $X$ so $\mbox{Im}\;p|_{U_\alpha}\cong U_{\alpha}\times\mathbb{C}^N$, and we have transition functions $g_{\alpha\beta}:U_\alpha\cap U_\beta\to \mbox{GL}(N;\mathbb{C})$, describing how the local frame fields glue together. The transition functions have to satisfy the cocycle condition over triple intersections:
\begin{align}
g_{\alpha\beta} g_{\beta \gamma}=g_{\alpha\gamma} \text{ over } U_\alpha\cap U_\beta \cap U_\gamma.
\end{align}
We can then build an open cover $\{U_{\alpha,+},U_{\beta,-}\}$ of $S^2\times X$, with $U_{\alpha,\pm }=U_{\alpha}\times D_{\pm}$, where $D_{\pm}$ are the two disks in $S^2=D_{+}\cup D_{-}$. The transition functions for the bundle $p_1^{*}\overline{\mathcal{L}}\otimes p_2^{*}\mbox{Im}\;p\oplus p_1^*\mathcal{L}\otimes p_2^*\theta^N$ are then of the form
\begin{itemize}
\item[(i)] over $U_{\alpha,-}\cap U_{\beta,+}$ the transition function $g_{\alpha -,\beta +}=\text{diag}(\frac{\bar{z}}{|z|}g_{\alpha\beta}, \frac{z}{|z|}I_N)$
\item[(ii)] over $U_{\alpha,\pm}\cap U_{\beta,\pm}$ the transition function $g_{\alpha \pm,\beta \pm}=\text{diag}(g_{\alpha\beta}, I_N)$.
\end{itemize}
We can define a path of transition functions as follows. Let
\begin{align}
R_{t}=\left[\begin{array}{cc}
\cos\left(\frac{\pi (1-t)}{2}\right) I_{N} & -\sin\left(\frac{\pi (1-t)}{2}\right)I_{N}\\
\sin\left(\frac{\pi (1-t)}{2}\right)I_{N} & \cos\left(\frac{\pi (1-t)}{2}\right)I_{N}
\end{array}\right]
\end{align}
and take
\begin{align}
&g_{\alpha-,\beta+}(t)=\left[\begin{array}{cc}
\frac{\bar{z}}{|z|}g_{\alpha\beta} & 0\\
0 & I_N
\end{array}
\right] R_t\left[\begin{array}{cc}
\frac{z}{|z|}I_N & 0\\
0 & I_N 
\end{array}
\right]R_{-t},\nonumber\\
&g_{\alpha-,\beta-}(t)=\left[\begin{array}{cc}
g_{\alpha\beta} & 0\\
0 & I_N
\end{array}\right],\nonumber\\
&g_{\alpha+,\beta+}(t)\nonumber\\
&=R_t\left[\begin{array}{cc}
\frac{\bar{z}}{|z|}I_N & 0\\
0 & I_N
\end{array}
\right]R_{-t}\left[\begin{array}{cc}
g_{\alpha\beta} & 0\\
0 & I_N
\end{array}\right]R_t\left[\begin{array}{cc}
\frac{z}{|z|}I_N & 0\\
0 & I_N
\end{array}
\right]R_{-t},
\end{align}
with $t\in[0,1]$. Then one can check that, indeed, for $\alpha,\beta,\gamma\in \{1,...,M\}$, we have the cocycle conditions
\begin{align}
g_{\alpha+,\beta+}g_{\beta+,\gamma+}&=g_{\alpha+,\gamma+}, \text{ over } U_{\alpha,+}\cap U_{\beta,+}\cap U_{\gamma, +},\nonumber\\
g_{\alpha-,\beta-}g_{\beta-,\gamma-}&=g_{\alpha-,\gamma-}, \text{ over } U_{\alpha,-}\cap U_{\beta,-}\cap U_{\gamma, -},\nonumber\\
g_{\alpha-,\beta+}g_{\beta+,\gamma+}&=g_{\alpha-,\gamma+}, \text{ over } U_{\alpha,-}\cap U_{\beta,+}\cap U_{\gamma, +},\nonumber\\
g_{\alpha-,\beta-}g_{\beta-,\gamma+}&=g_{\alpha+,\gamma+}, \text{ over } U_{\alpha,-}\cap U_{\beta,-}\cap U_{\gamma, +},
\end{align}
needed to define a vector bundle over $[0,1]\times (S^2\times X)$. Observe that the resulting bundle restricted to $\{0\}\times S^2\times X$ is $p_1^{*}\overline{\mathcal{L}}\otimes p_2^{*}\mbox{Im}\;p\oplus p_1^*\mathcal{L}\otimes p_2^*\theta^N$ and the one restricted to $\{1\}\times S^2\times X$ is $p_2^{*}\left(\mbox{Im}\;p\oplus \theta^{N}\right)$. By homotopy invariance of topological vector bundles, the result of Eq.~\eqref{eq:a useful isomorphism} follows. Plugging this result in Eq.~\eqref{eq: E stable iso to part 1}, we get
\begin{widetext}
\begin{align}
&p_1^{*}\mathcal{L} \otimes p_2^{*}\mbox{Im}\;q \oplus p_1^{*}\overline{\mathcal{L}}\otimes p_2^{*}\mbox{Im}\;p\oplus p_1^*\mathcal{L}\otimes p_2^*\theta^N\oplus p_1^*\mathcal{L}^{\perp}\otimes p_2^*\theta^N\nonumber\\
&\cong p_1^{*}\mathcal{L} \otimes p_2^{*}\mbox{Im}\;q \oplus p_2^{*}\left(\mbox{Im}\;p\oplus \theta^N\right)\oplus p_1^*\mathcal{L}^{\perp}\otimes p_2^*\theta^N\nonumber\\
&\cong p_1^{*}\mathcal{L} \otimes p_2^{*}\mbox{Im}\;q \oplus p_1^{*}\theta^1\otimes p_2^{*}\mbox{Im}\;p\oplus p_1^*\mathcal{L}^{\perp}\otimes p_2^*\theta^N\oplus p_2^*\theta^N\nonumber\\
&\cong p_1^{*}\mathcal{L} \otimes p_2^{*}\mbox{Im}\;q \oplus p_1^{*}\theta^1\otimes p_2^{*}\left(\mbox{Im}\;q\right)^\perp\oplus p_1^*\mathcal{L}^{\perp}\otimes p_2^*\theta^N\oplus \theta^N\nonumber\\
&\sim_{s} p_1^*\mathcal{L} * p_2^*\mbox{Im}\;q.
\end{align}
\end{widetext}
Observe that what we have proved is that 
\begin{align}
[p_1^{*}\mathcal{L} \otimes p_2^{*}\mbox{Im}\;q \oplus p_1^{*}\overline{\mathcal{L}}\otimes p_2^{*}\mbox{Im}\;p]_s=[p_1^*\mathcal{L} * p_2^*\mbox{Im}\;q]_s,
\end{align}
and we have equivalent results in terms of stable equivalence classes of projectors and Hamiltonians. In particular, it yields that the generalized Dirac monopole built out of $\{h(x)\}_{x\in X}$ of Eq.~\eqref{eq: generalized Dirac monopole} is equivalent to the (external tensor) product of the Dirac monopole of Eq.~\eqref{eq: Dirac monopole} by $\{h(x)\}_{x\in X}$. 

Bott periodicity in complex \emph{K}-theory is a consequence of the observation that the external tensor product of vector bundles, that takes vector bundles $E\to X$ and $F\to Y$ to $p_1^* E\otimes p_2^* F$, where $p_1,p_2$ are the natural projections, induces an isomorphism $K^0(S^2\times X)\cong K^0(X)$. We remark that, as a ring, $K^{0}(S^2)=\mathbb{Z}[x]/(x-1)^2$, where $x=[\mathcal{L}]\in K^0(S^2)$. At the level of reduced \emph{K}-theory, the Bott periodicity isomorphism $\widetilde{K}^0(X)\cong \widetilde{K}(S^2\wedge X)=\widetilde{K}^{-2}(X)$, consists precisely in multiplication by the \emph{Bott} class $[\mathcal{L}]-[1]\leftrightarrow [\mathcal{L}]_s$, which is exactly the non-trivial $S^2$ piece appearing in the external tensor product $[E]_s=[p_1^{*}\mathcal{L} * p_2^*\mbox{Im}\;q]_s$ constructed above.
Observe that we have done this using the $+1$ eigenbundle, but the analogous result holds for the orthogonal complement bundle since their sum is trivial, namely, we would get a bundle isomorphic to $p_1^{*}\mathcal{L} * p_2^*\mbox{Im}\;p$.
\section{Second Chern number for the external tensor product of Dirac monopoles}
\label{sec:2nd chern number for the external tensor product of dirac monopoles}
We wish to compute the second Chern number associated with the positive energy eigenbundle of the Hamiltonian
\begin{align}
\widetilde{H}(y)=\sum_{i=1}^{5}y^i\gamma_i,\  y\in S^4,
\end{align}
representing the stable equivalence class of the external tensor product of two Dirac monopoles Eq.~\eqref{eq: external tensor product of two dirac monopoles}. The positive energy eigenbundle is described by the orthogonal projector $P(y)=(I_4+H(y))/2$. The associated Berry curvature is given by the matrix valued two-form
\begin{align}
F= P dP\wedge dP P&= \frac{1}{4}P\sum_{i,j=1}^{5}\gamma_i\gamma_j P dy^i\wedge dy^j \nonumber\\
&=\frac{1}{8}P\sum_{i,j=1}^{5}[\gamma_i,\gamma_j] P dy^i\wedge dy^j.
\end{align}
The second Chern class is then represented by the closed differential form
\begin{align}
&\frac{1}{8\pi^2}\left(\tr\; F^2 - \tr\; F\wedge \tr \; F\right)\nonumber\\
&=\frac{1}{8\pi^2}\tr\; F^2 \nonumber\\
&=\frac{1}{8\pi^2}\frac{1}{16}\sum_{i,j,k,l=1}^{5} \tr\;\left(P\gamma_i\gamma_j\gamma_k\gamma_l \right) dy^{i}\wedge dy^j \wedge dy^{k}\wedge dy^{l},
\end{align}
where we used the fact that the trace of a commutator is zero and the cyclic property of the trace. Now, if we write $P(y)=(I_4+H(y))/2$, the first term does not contribute. The reason is that from the collection of $5$ gamma matrices, this term will be missing one, which, in turn, anti-commutes with the $4$ and squares to the identity. We are then left with
\begin{align}
\frac{1}{8\pi^2}\frac{1}{32}\sum_{i,j,k,l,m=1}^{5} \tr\;\left(\gamma_i\gamma_j\gamma_k\gamma_l\gamma_m \right) y^{i}dy^{j}\wedge dy^k \wedge dy^{l}\wedge dy^{m}.
\end{align}
In the above sum, if $\gamma_i$ is equal to any to any of the other $\gamma$'s appearing in the product the trace yields zero by the same argument as before. Therefore, we are left with the completely anti-symmetric combination,
\begin{align}
-\frac{1}{8\pi^2}\frac{1}{8}\sum_{i,j,k,l,m=1}^{5} \varepsilon_{ijklm} y^{i}dy^{j}\wedge dy^k \wedge dy^{l}\wedge dy^{m},
\end{align}
where $\varepsilon_{ijklm}$ is the Levi-Civita symbol, and where we noted that $\tr\;\left(\gamma_1\gamma_2\gamma_3\gamma_4\gamma_5 \right)=-\tr \; \gamma_5^2=-4$. Finally observe that the differential form $(1/4!)\sum_{i,j,k,l,m=1}^{5} \varepsilon_{ijklm} y^{i}dy^{j}\wedge dy^k \wedge dy^{l}\wedge dy^{m}$ is nothing but the volume element of $S^4$ according to the standard round metric, which integrates to $8\pi^2/3$. Upon integrating over $S^4$, we obtain the result $-1$.
%%%%%
%\bibliographystyle{unsrt}
\bibliography{bib}
\end{document}